\documentclass[twocolumn,superscriptaddress,amsmath,amssymb,aps,longbibliography,prl,floatfix]{revtex4}
\usepackage{txfonts}
\usepackage{amsfonts}
\usepackage{amsmath}
\usepackage{graphicx}
\usepackage{diagbox}
\usepackage{amssymb}
\usepackage{bbm}
\usepackage{subfigure}
\usepackage{url}
\usepackage{multirow}
\usepackage{bm}
\usepackage{booktabs}
\usepackage{array}
\usepackage{color}
\usepackage{algorithm}
\usepackage{algorithmic}
\makeatletter
\edef\ftype@algorithm{\the\c@float@type}
\newcommand\floatc@algleft[2]{%
  \setbox\@tempboxa\hbox{{\@fs@cfont #1} #2}%
  \ifdim\wd\@tempboxa>\hsize
    {\@fs@cfont #1} #2\par
  \else
    \hbox to\hsize{\box\@tempboxa\hfil}%
  \fi
}
\newcommand\fs@algleft{%
  \def\@fs@cfont{\bfseries}\let\@fs@capt\floatc@algleft
  \def\@fs@pre{\hrule height.8pt depth0pt \kern2pt}%
  \def\@fs@post{\kern2pt\hrule\relax}%
  \def\@fs@mid{\kern2pt\hrule\kern2pt}%
  \let\@fs@iftopcapt\iftrue
}
\floatstyle{algleft}\restylefloat{algorithm}
\makeatother

\usepackage{pifont}
\newcommand{\cmark}{\textcolor{tabgreen}{\ding{51}}}
\newcommand{\xmark}{\textcolor{tabred}{\ding{55}}}

\usepackage{makecell}
\newcommand{\PreserveBackslash}[1]{\let\temp=\\#1\let\\=\temp}
\newcolumntype{C}[1]{>{\PreserveBackslash\centering}p{#1}}
\newcolumntype{R}[1]{>{\PreserveBackslash\raggedleft}p{#1}}
\newcolumntype{L}[1]{>{\PreserveBackslash\raggedright}p{#1}}

\usepackage{xcolor}
\definecolor{tabred}{RGB}{238, 32, 36}
\definecolor{tabblue}{RGB}{67, 120, 188}
\definecolor{tabgreen}{RGB}{0, 153, 68}

\usepackage[bookmarks=true,colorlinks,citecolor=blue,urlcolor=blue]{hyperref}
\makeatletter
\providecommand*{\toclevel@algorithm}{0}
\makeatother

\begin{document}

\title{Ab initio simulation of the first-order proton-ordering transition in water ice}

\author{Qi Zhang}
\affiliation{Beijing National Laboratory for Condensed Matter Physics and Institute of Physics, Chinese Academy of Sciences, Beijing 100190, China}

\author{Sicong Wan}
\affiliation{School of Physics, Renmin University of China, Beijing 100872, China}

\author{Lei Wang}
\email{wanglei@iphy.ac.cn}
\affiliation{Beijing National Laboratory for Condensed Matter Physics and Institute of Physics, 
Chinese Academy of Sciences, Beijing 100190, China}

\date{\today}
	
\begin{abstract}

Proton ordering in water ice is a paradigmatic order-disorder transition in a locally constrained system.
The ice rules require exactly two hydrogens close to each oxygen, restricting the disorder to an exponentially large yet strongly correlated manifold of hydrogen-bond configurations.
Within this constrained space, meV-scale energy differences drive the transition from disordered ice Ih to ordered ice XI, while distinct configurations are separated by eV-scale barriers. 
These barriers hinder equilibration in experiments, and efficient sampling of this space with the required energy accuracy has remained a long-standing challenge in simulation. 
We address this by combining a machine learning interatomic potential with loop updates that preserve the ice rules and continuous updates of atomic coordinates, enabling equilibrium sampling with \emph{ab initio} accuracy and capturing configurational entropic effects.
In systems of up to $360$ water molecules, with over $10^6$ samples retained per temperature point, the simulations reveal clear first-order transition signatures at $83~\mathrm{K}$: a negative Binder cumulant, a bimodal potential energy distribution, and a sharp step in the lattice aspect ratio.  
Nuclear quantum effects are estimated to lower the transition temperature by approximately $20~\mathrm{K}$, bringing the prediction closer to the experimental value of $72~\mathrm{K}$.

\end{abstract}
\maketitle


In the most common form of solid water, ice Ih, oxygen atoms form a hexagonal lattice, while hydrogen atoms remain disordered but strongly correlated by the ice rules, which require each oxygen atom to have two nearby hydrogens and two more distant ones~\cite{keen2015thecrystallography}.
Pauling first estimated that the number of ice-rule configurations for $N$ water molecules scales approximately as $1.5^{N}$~\cite{pauling1935structure, macdowell2004combinatorial, herrero2013configurational, berg2007residual, xu2025equivalence}, implying a finite residual configurational entropy at low temperature.
Ice Ih is thus a canonical example of macroscopic degeneracy arising from local constraints.
Because configurations that violate the ice rules correspond to high-energy defect states, they are strongly suppressed.
Electrostatic interactions lift the degeneracy among the allowed configurations by only a few meV per molecule~\cite{tribello2006proton}, but eV-scale barriers separate distinct hydrogen-bond configurations, making proton reordering kinetically hindered~\cite{koning2006orientational, lin2011correlated}.
Under natural conditions, the transformation from disordered ice Ih to ordered ice XI can require $10^4$--$10^5$ years~\cite{fukazawa2006ferroelectric,kobayashi2012icxi,yen2015proton}.
KOH doping introduces ionic defects that accelerate the transition to a timescale of days and reveals a first-order calorimetric signal around $72$--$73~\mathrm{K}$~\cite{kawada1972dielectric, tajima1982phase, tajima1984calorimetric, yen2015proton}, but the observed configurational entropy loss is sensitive to both the dopant type and the cooling timescale.

Unlike magnetic spin ices~\cite{ramirez1999zero, denhertog2000dipolar, melko2001longrange}, which obey the same ice rules but are effectively described by only discrete Ising variables, water ice couples continuous atomic coordinates to a discrete hydrogen-bond network.
Resolving competing proton-ordering states also demands high energetic accuracy, challenging first-principles simulations.
As summarized in the Supplementary Information (SI), these challenges have prevented previous studies from fully resolving the transition~\cite{barkema1993properties,rick2003dielectric,singer2005hbtopology,knight2006hbtopology,schoenherr2014dielectric,piaggi2021enhancing,zhang2021phase}.
Empirical potentials such as the transferable intermolecular potential with 4 points (TIP4P) and its variants~\cite{jorgensen1983tip4p, rick1994tip4pfq} are efficient but yield inconsistent energetic rankings~\cite{rick2003dielectric, hirsch2004quantum, raza2011proton}.
They also fail to reproduce the ferroelectric $Cmc2_1$ ground state observed in density functional theory (DFT) calculations and experiments~\cite{rick2003dielectric, hirsch2004quantum, raza2011proton, leadbetter1985equilibrium, howe1989adetermination, jackson1997single}.
Graph-invariant models improve the ordering of distinct hydrogen-bond topologies but omit atomic vibrations and cell fluctuations, introducing uncontrolled biases~\cite{singer2005hbtopology, knight2006hbtopology}.
Monte Carlo sampling with on-the-fly DFT energy evaluations is prohibitively expensive, restricting system sizes and sampling depth~\cite{schoenherr2014dielectric}.
Although machine-learning interatomic potentials (MLIPs) enable large-scale sampling at high accuracy, existing MLIPs still struggle to correctly rank hydrogen-bond configurations and stabilize the true ground state~\cite{piaggi2021enhancing, zhang2021phase}.

To accurately describe the phase transition between water ice Ih and XI, one needs to
(i) resolve meV-per-molecule energy splittings among competing hydrogen-bond configurations, and
(ii) sample both configurational and vibrational degrees of freedom on equal footing.
In this work, we train an MLIP on a DFT dataset spanning a broad range of hydrogen-bond configurations to retain \emph{ab initio} accuracy for the potential energies. 
Then, we devise a Monte Carlo sampling scheme that combines discrete loop updates~\cite{rahman1972proton, rick2003dielectric} that preserve the ice rules with the Metropolis-adjusted Langevin algorithm (MALA)~\cite{roberts1996exponential} for atomic coordinate updates, thereby enabling ergodic sampling of hydrogen-bond configurations while capturing atomic vibrations and lattice deformations.
In this approach, configurational entropy is sampled directly rather than treated through a separate approximation.
Together with further performance optimizations of MLIP inference, these advances make large-scale, high-statistics simulations of proton ordering in water ice feasible, enabling controlled finite-size analyses of the ice Ih to XI transition.

\vspace{1em}
\textbf{Potential energy surface}

Hydrogen-bond configurations differ in energy by only a few $\mathrm{meV}$ per molecule, making the choice of exchange-correlation functional critical for resolving proton ordering.
Diffusion Monte Carlo benchmarks~\cite{dellapia2022dmcice13} indicate that the strongly constrained and appropriately normed (SCAN)~\cite{sun2015scan} functional and its regularized variant $\mathrm{r^2SCAN}$~\cite{furness2020r2scan} systematically improve on Perdew-Burke-Ernzerhof (PBE)~\cite{perdew1996generalized} and remain competitive with widely used dispersion-corrected approaches~\cite{grimme2006semiempirical} for ice energetics across multiple phases.
Hybrid functionals such as PBE0~\cite{adamo1999toward} offer comparable energetic accuracy but increase computational cost by one to two orders of magnitude.
We therefore adopt $\mathrm{r^2SCAN}$ in this work for its greater numerical stability than SCAN.
To balance accuracy and efficiency, we employ a high-order equivariant message-passing neural network, namely the MACE model~\cite{batatia2022mace,batatia2025design}.
Equivariance and message passing are both critical for capturing these subtle energy differences among competing hydrogen-bond topologies.
We train the model on more than $40{,}000$ DFT-$\mathrm{r^2SCAN}$ configurations spanning a wide range of hydrogen-bond topologies.
The model achieves test-set root-mean-square errors (RMSEs) of $0.26~\mathrm{meV}/\mathrm{H_2O}$ for energies and $2.9~\mathrm{meV}/\mathrm{\AA}$ for forces.
Additional details are provided in the Methods section and SI~\cite{supp}.

\begin{figure}[ht]
\centering
\includegraphics[width=1.0\columnwidth]{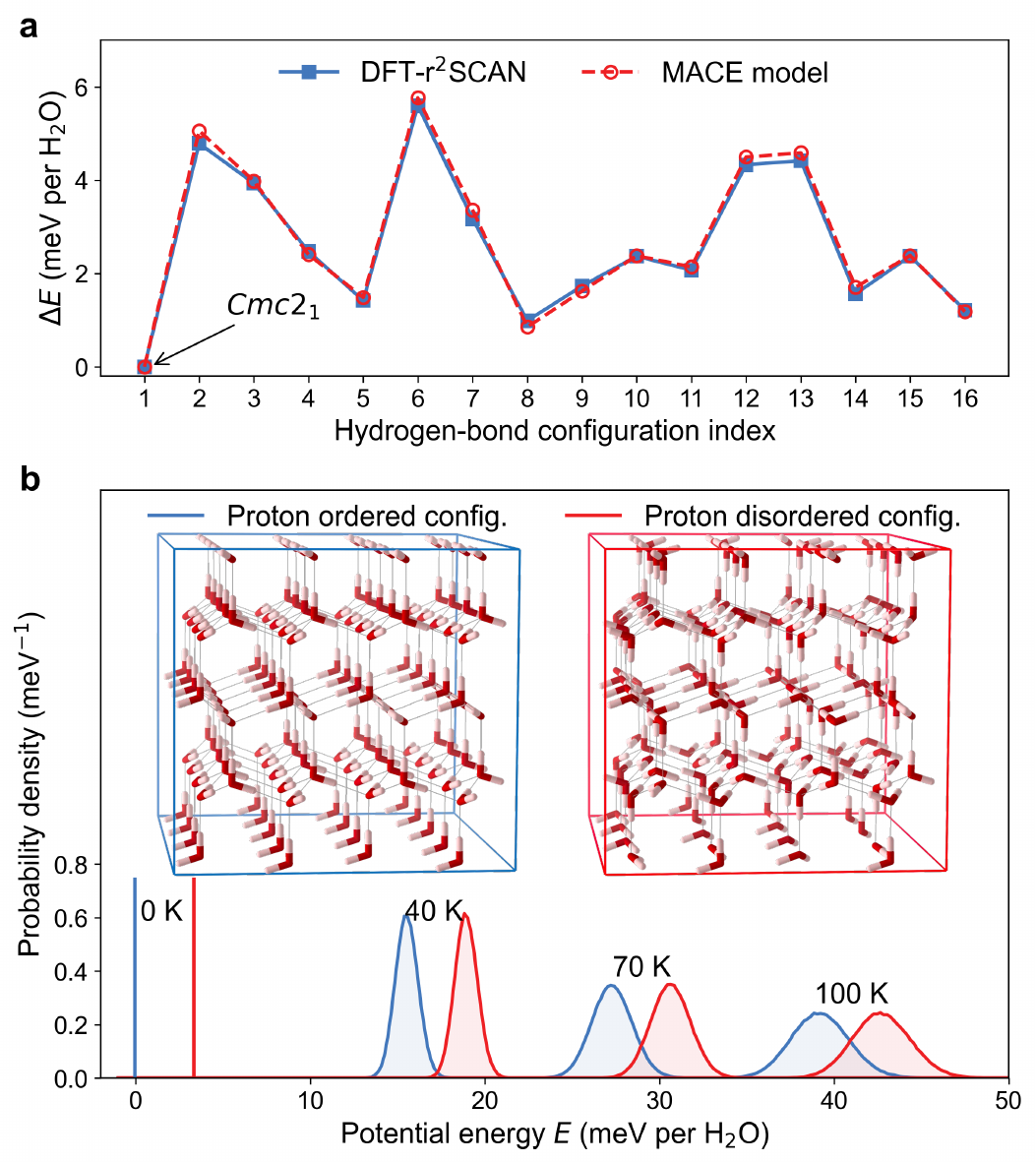}
\caption{ 
\textbf{Relative potential energies of hydrogen-bond configurations.}
\textbf{a}, Potential energies of all $16$ symmetry-inequivalent hydrogen-bond configurations in a cell containing $N=8$ water molecules, computed with DFT-$\mathrm{r^2SCAN}$ and the MACE potential.
Energies are referenced to the $Cmc2_1$ configuration, $\Delta E = E_i - E_1$.
\textbf{b}, Potential energy distributions at different temperatures for an $N=128$ cell, comparing the proton-ordered $Cmc2_1$ configuration with a random proton-disordered configuration.
Atomic structures were visualized using VESTA software~\cite{momma2011vesta} with oxygen in red and hydrogen in pale pink.
}
\label{fig:dft}
\end{figure}

To assess the accuracy of the MACE model for energies within the ice-rule manifold, we consider the same cell with $N=8$ molecules as in Ref.~\cite{hirsch2004quantum} and evaluate all $16$ symmetry-inequivalent hydrogen-bond configurations among the $114$ ice-rule states.
As shown in Fig.~\ref{fig:dft}\textbf{a}, MACE is in excellent agreement with the $\mathrm{r^2SCAN}$ reference, with absolute errors of $0.1$--$0.2~\mathrm{meV}/\mathrm{H_2O}$.
The model correctly identifies the ferroelectric $Cmc2_1$ structure as the lowest energy state and reproduces the energy ordering across all $16$ configurations.
Notably, neither the configurations shown here nor any other configurations from the $N=8$ cell were included in the training set, demonstrating robust generalization to unseen structures.
By contrast, commonly used empirical force fields~\cite{rick2003dielectric} and some non-equivariant MLIPs~\cite{piaggi2021enhancing,zhang2021phase} can fail to reproduce these meV-scale energy splittings or identify the correct ground state.

To quantify the impact of atomic vibrational degrees of freedom, we sample finite-temperature atomic fluctuations using a MALA-based Monte Carlo scheme~\cite{roberts1996exponential}.
In this procedure, both oxygen and hydrogen atoms are displaced while the hydrogen-bond topology is kept fixed.
As shown in Fig.~\ref{fig:dft}\textbf{b}, we consider an $N=128$ supercell and compare a fully proton-ordered ferroelectric $Cmc2_1$ configuration with a randomly generated proton-disordered configuration.
The two vertical lines mark the zero-temperature relaxed energies of the corresponding structures, separated by $3.40~\mathrm{meV}/\mathrm{H_2O}$.
Thermal fluctuations increase the mean energies and broaden the energy distributions.
At the low temperature of $40~\mathrm{K}$, a slight overlap between the two distributions emerges.
The overlap grows with temperature because the distributions broaden substantially, even though the mean energy separation increases slightly to $3.52~\mathrm{meV}/\mathrm{H_2O}$ at $100~\mathrm{K}$.
This demonstrates that relying solely on zero-temperature relaxed energies can lead to uncontrolled errors in predicting the Ih to XI transition~\cite{singer2005hbtopology,knight2006hbtopology}, as finite-temperature vibrations renormalize the effective energetic competition among hydrogen-bond configurations.

\vspace{1em}
\textbf{Composite Monte Carlo sampling}

A central challenge in computational studies of the proton ordering transition is that transitions between distinct hydrogen-bond configurations require moving through eV-scale energy barriers, far exceeding thermal energies at relevant temperatures.
Consequently, continuous coordinate-space moves in standard molecular dynamics simulations fail to connect distinct hydrogen-bond topologies, and the dynamics become kinetically trapped within a single hydrogen-bond configuration.
Any local attempt to reorient an individual water molecule violates the ice-rule constraint that each oxygen must have two nearby and two distant hydrogens, thereby creating high-energy defects.

To enable efficient sampling of ice-rule configurations, we combine Monte Carlo updates of discrete and continuous degrees of freedom while maintaining detailed balance and ergodicity.
Loop moves update the discrete hydrogen-bond topology using random-walk short-loop proposals~\cite{rahman1972proton, rick2003dielectric}.
The proposals construct closed loops on the hydrogen-bond network and reverse bond directions while preserving the ice rules.
In particular, winding loops traverse periodic boundaries, thereby connecting topologically distinct sectors and enabling transitions between configurations with different polarization magnitudes.
We validate this approach by exhaustive enumeration in a system of $N=16$ molecules, confirming that short-loop updates can access all $2{,}970$ ice-rule configurations.
The move set is therefore ergodic in principle within the ice-rule manifold.
Continuous degrees of freedom are sampled at fixed topology using MALA and cell moves.
MALA updates atomic coordinates using force-guided proposals~\cite{roberts1996exponential}, with forces computed efficiently via automatic differentiation of the MACE potential.
Cell moves update the lattice parameters to sample volume fluctuations~\cite{frenkel2023understanding}.
Full algorithmic details are provided in SI~\cite{supp}.

We perform Monte Carlo sampling in the isothermal--isobaric ensemble at zero pressure across $T=20$--$200~\mathrm{K}$ for supercells containing $N=64$--$360$ water molecules.
For the largest system size, each temperature point involves more than $10^8$ Monte Carlo proposals, with nearly $2\times 10^6$ configurations retained after thinning.
To sustain this sampling throughput, we optimize graph construction for MACE evaluations, yielding a speedup approaching an order of magnitude in the overall simulation workflow~\cite{supp}.
Even with these optimizations, the full simulation campaign required approximately $3\times 10^4$ graphics processing unit (GPU) hours.
Loop moves are nonlocal updates that move many hydrogen atoms simultaneously, generating large collective changes in hydrogen-bond configurations and consequently low acceptance rates.
At $100~\mathrm{K}$, the loop acceptance rate is $1.6\%$. 
It falls below $0.01\%$ at $75~\mathrm{K}$ and decreases by further orders of magnitude upon cooling~\cite{supp}.
By contrast, after step-length tuning, the acceptance rates for continuous updates, including MALA and cell moves, remain between $40\%$ and $70\%$.

\vspace{1em}
\textbf{First-order phase transition signatures}

\begin{figure*}[ht]
\centering
\includegraphics[width=2.0\columnwidth]{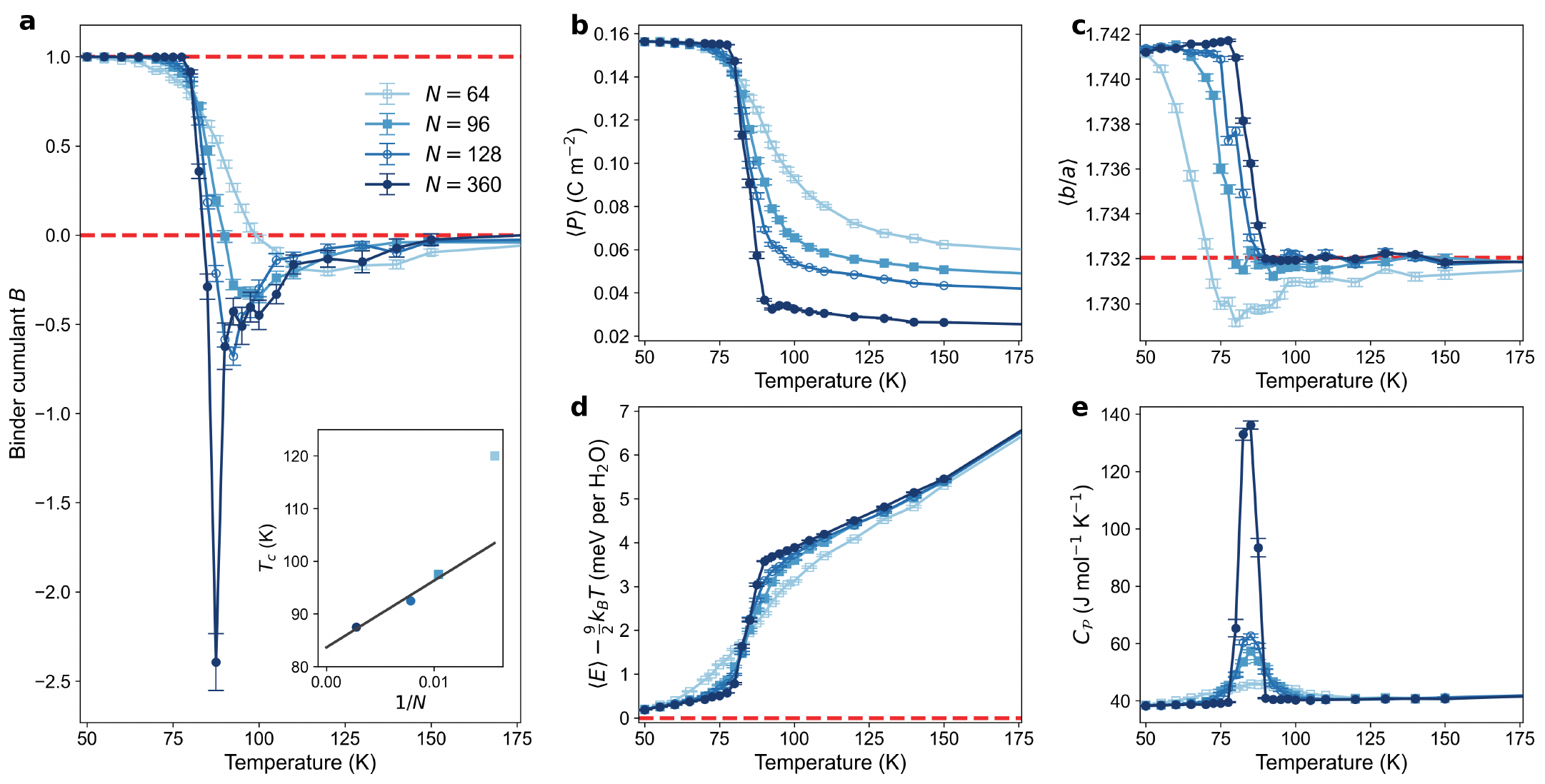}
\caption{
\textbf{Thermodynamic and structural results of the proton ordering transition.}
\textbf{a}, Binder cumulant $B$ of the total dipole moment magnitude $M$ for different system sizes $N$ (number of $\mathrm{H_2O}$ molecules).
The inset shows the temperature at which $B$ reaches its minimum for each $N$, which is extrapolated to the thermodynamic limit. 
\textbf{b}, Polarization density magnitude $\langle P \rangle$.
\textbf{c}, Orthorhombic lattice anisotropy ratio $\langle b/a\rangle$ of the unit cell.
The dashed line marks the ideal hexagonal close-packed value $\sqrt{3}$.
\textbf{d}, Potential energy per molecule $\langle E \rangle$ with the classical harmonic contributions $\frac{9}{2}k_{\mathrm{B}}T$ (dashed line) subtracted, isolating the anharmonic and hydrogen-bond configurational contributions.
\textbf{e}, Molar heat capacity $C_{\mathcal{P}}$ estimated from the enthalpy fluctuations excluding kinetic contributions, including vibrational and hydrogen-bond configurational contributions.
}
\label{fig:results}
\end{figure*}

To quantify proton ordering, we use the total dipole moment $\bm{M}$ as an order parameter and compute the Binder cumulant $B$~\cite{binder1984finite} to characterize its fluctuations (see Methods).
We adopt a normalized definition of $B$ such that $B\to 1$ in the ordered phase and $B\to 0$ in the disordered phase.
For a continuous transition, the curves for different system sizes intersect near the transition temperature, whereas a first-order transition yields a negative dip that deepens with increasing system size~\cite{binder1984finite,challa1986finite}.
In Fig.~\ref{fig:results}\textbf{a}, $B$ exhibits the expected limiting behavior at both low and high temperatures.
For $N=64$, the negative minimum is shallow and broad, whereas it deepens and sharpens markedly with increasing system size, becoming pronounced at $N=360$.
This size dependence provides clear evidence for a first-order transition, as expected from two-phase coexistence near the transition and the resulting multimodal distribution of the polarization magnitude $P$.
Accordingly, we define the finite-size transition temperature $T_{\mathrm{c}}$ as the temperature at which $B$ reaches its minimum. 
The inset shows the thermodynamic limit extrapolation, yielding a transition temperature of $83~\mathrm{K}$.

We characterize ferroelectric order using the polarization density magnitude $P$, defined as the volume-normalized dipole moment (see Methods).
We use $P$ as a structural proxy for molecular orientation, but its absolute value is not directly comparable to experiments or DFT because it is computed from the atomic coordinates and neglects electronic polarization and higher-order multipoles.
As shown in Fig.~\ref{fig:results}\textbf{b}, it saturates near $0.16~\mathrm{C\,m^{-2}}$ at low temperatures for all system sizes, and the corresponding configurations are consistent with the ferroelectric $Cmc2_1$ structure reported in experiments and DFT~\cite{rick2003dielectric, hirsch2004quantum, raza2011proton, leadbetter1985equilibrium, howe1989adetermination, jackson1997single}.
Upon warming, the polarization decreases gradually and then drops sharply around $80~\mathrm{K}$.
The drop sharpens with increasing system size, consistent with the first-order character of the transition.
At high temperatures, $P$ is only weakly temperature dependent and decreases with system size, consistent with an approach to an isotropic, nonpolar state in the thermodynamic limit. 
Accordingly, the residual high-temperature polarization is attributable to finite-size effects rather than true long-range order.

We monitor the lattice response to the phase transition through the orthorhombic anisotropy ratio $\langle b/a\rangle$, shown in Fig.~\ref{fig:results}\textbf{c}.
The lattice exhibits a small but systematic anisotropic distortion associated with ferroelectric ordering.
Accordingly, $\langle b/a\rangle$ is approximately $1.741$ in the ordered phase ice XI, and upon warming it approaches the ideal hexagonal value $\sqrt{3}$ expected for proton-disordered ice Ih.
This trend agrees with high-resolution neutron powder diffraction experiments, which found that ice XI exhibits a larger $b/a$ ratio than ice Ih~\cite{line1996ahigh, biggs2024proposing}.
Notably, $\langle b/a\rangle$ shows a sharp step around $85~\mathrm{K}$, with the apparent jump shifting slightly to higher temperatures as the system size $N$ increases.
This abrupt change and its systematic finite-size evolution therefore provide a structural signature of the order--disorder transition.
To resolve the origin of the anisotropy change, we examine the individual lattice parameters.
Within the narrow transition region on warming from ice XI to ice Ih, $b$ contracts while $a$ and $c$ elongate (see SI~\cite{supp}).
A similar behavior is also observed in the anisotropy ratio $\langle c/a\rangle$~\cite{supp}.

The potential energy $E$ is dominated by lattice vibrations, with the harmonic term contributing $\frac{9}{2}\,k_{\mathrm{B}}T$ per molecule and varying by several tens of meV across the temperature range studied~\cite{supp}.
Figure~\ref{fig:results}\textbf{d} shows the values after subtracting this harmonic baseline, leaving the anharmonic vibrational contribution and variations among hydrogen-bond configurations.
For $N=360$, the system remains predominantly in the $Cmc2_1$ configuration below $75~\mathrm{K}$, where the anharmonic contribution is about $1~\mathrm{meV/H_2O}$ and increases gradually with temperature.
Around $80~\mathrm{K}$, the potential energy exhibits a size-dependent jump that sharpens with size, reaching $3~\mathrm{meV/H_2O}$, consistent with hydrogen-bond reordering across the transition.
Although the potential energy increases, the number of accessible hydrogen-bond configurations grows rapidly in this temperature range, suggesting a gain in configurational entropy that compensates the energetic cost and favors the disordered phase.
The systematic sharpening of the jump with increasing system size indicates a discontinuity in the thermodynamic limit, providing further evidence for a first-order transition.
Across all system sizes, the energy-curve crossing temperatures cluster around $82$--$83~\mathrm{K}$, in line with the Binder-cumulant estimate.

We compute the molar heat capacity $C_{\mathcal{P}}$ from the enthalpy fluctuations excluding kinetic contributions (see Methods).
As shown in Fig.~\ref{fig:results}\textbf{e}, the heat capacity contains contributions from lattice vibrations and hydrogen-bond configurational fluctuations.
In our classical simulations, lattice vibrations contribute a nearly temperature-independent background of approximately $40~\mathrm{J\,mol^{-1}\,K^{-1}}$.
This large vibrational background makes the order--disorder signal difficult to detect, demanding extensive sampling.
For smaller systems ($N=64$ and $96$), the heat capacity exhibits only a broad, gentle hump near the transition.
With increasing system size, this hump sharpens and evolves into a pronounced peak.
For $N=360$, the peak is clear, narrow, and well-defined, which is characteristic of a first-order transition.
Such a narrow peak contrasts with the broad critical anomaly characteristic of continuous transitions.
The heat-capacity peak occurs at $85~\mathrm{K}$ for all system sizes, indicating that the transition temperature has essentially converged to its thermodynamic-limit value.
In addition, our estimate of the hydrogen-bond configurational entropy shows that the transition removes nearly all of Pauling's residual entropy, as shown in SI~\cite{supp}.

\vspace{1em}
\textbf{Energy bimodality and phase coexistence}

\begin{figure*}[ht]
\centering
\includegraphics[width=2.0\columnwidth]{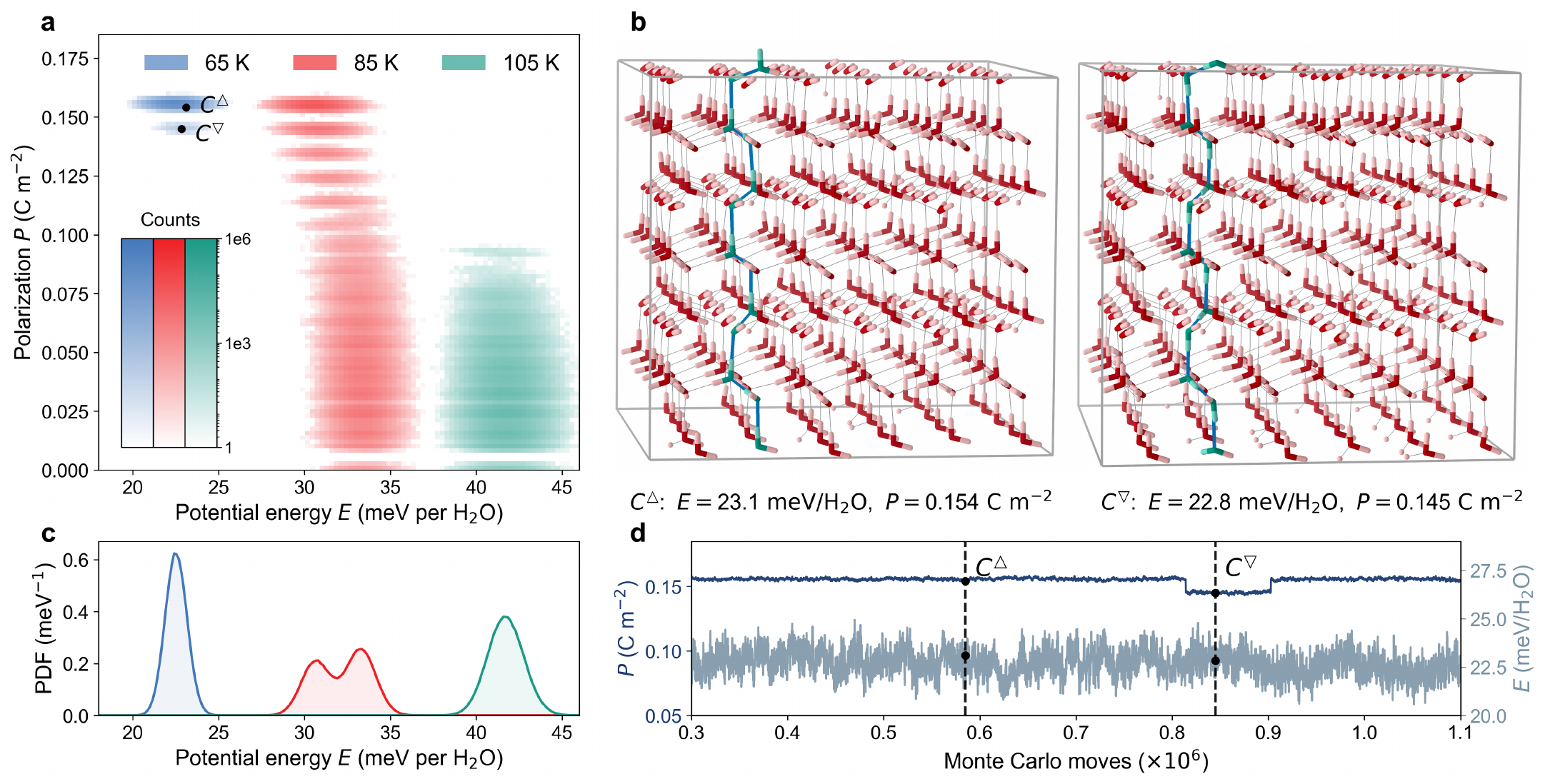}
\caption{
\textbf{Potential energy and polarization distributions with representative configurations.}
(a) Two-dimensional probability density of polarization magnitude $P$ versus potential energy per molecule $E$ from $10^6$ Monte Carlo samples at $65$, $85$, and $105$~K for an $N=360$ supercell.
Black markers highlight two selected configurations, labeled $C^{\bigtriangleup}$ and $C^{\bigtriangledown}$.
(b) Structural snapshots corresponding to the marked points.
Oxygen (dark teal) and hydrogen (light teal) atoms in the winding loop are highlighted.
A loop update flips the hydrogen atoms along the loop, transforming $C^{\bigtriangleup}$ into $C^{\bigtriangledown}$.
(c) Probability density function (PDF) of potential energy $E$ at the three temperatures.
(d) Monte Carlo trajectories at $65$~K for polarization magnitude $P$ (left axis) and potential energy $E$ (right axis).
Dashed vertical lines indicate configurations $C^{\bigtriangleup}$ and $C^{\bigtriangledown}$.
}
\label{fig:hist}
\end{figure*}

To better resolve phase coexistence, we analyze the joint distribution of the potential energy $E$ and the polarization density magnitude $P$.
For an $N=360$ supercell, we record $10^{6}$ configurations at $65$, $85$, and $105~\mathrm{K}$, as depicted in Fig.~\ref{fig:hist}\textbf{a}.
At $65~\mathrm{K}$ in the ice XI phase, $99.9\%$ of configurations fall into a higher-polarization cluster around $0.155~\mathrm{C\,m^{-2}}$, whereas the remaining $0.1\%$ form a rare lower-polarization cluster near $0.145~\mathrm{C\,m^{-2}}$.
We select representative configurations from these two clusters, labeled $C^{\bigtriangleup}$ and $C^{\bigtriangledown}$.
Here, $C^{\bigtriangleup}$ corresponds to the ferroelectric $Cmc2_1$ structure, whereas $C^{\bigtriangledown}$ can be obtained from $C^{\bigtriangleup}$ by a single winding-loop update that crosses the periodic boundary (Fig.~\ref{fig:hist}\textbf{b}).
The lower-polarization cluster lies entirely within the energy range of the higher-polarization one.
As shown in Fig.~\ref{fig:hist}\textbf{c}, despite the polarization contrast, the potential energy distribution at $65~\mathrm{K}$ remains unimodal, indicating substantial overlap.
Moreover, the acceptance rate for loop updates at this temperature is extremely low, on the order of $10^{-6}$~\cite{supp}.
A representative trajectory (Fig.~\ref{fig:hist}\textbf{d}) shows that the system spends most of the time in the $C^{\bigtriangleup}$ hydrogen-bond configuration, while rare winding-loop updates produce short-lived excursions to $C^{\bigtriangledown}$ that rapidly revert.
These loop updates are clearly visible in the polarization but leave little signature in the potential energy.

At $85~\mathrm{K}$, the energy distribution shifts upward and becomes clearly bimodal (Fig.~\ref{fig:hist}\textbf{a},\textbf{c}), suggesting phase coexistence near a first-order transition.
This bimodality is barely resolved in small supercells ($N=64$, $96$, and $128$), underscoring the need for large system sizes.
In this regime, a substantial population of low-polarization configurations emerges alongside the nearly ferroelectric states that dominant at low temperature, indicating the emergence of proton-disordered states.
Accordingly, ordered and disordered configurations coexist. 
In the joint $P$-$E$ distribution, high-polarization configurations correspond to proton-ordered states, whereas low-polarization configurations are predominantly disordered and have a slightly higher mean energy despite substantial energy overlap.
At $105~\mathrm{K}$, the energy distribution shifts upward and broadens, returning to a single-peak shape.
Nearly all samples fall in the region $P<0.10~\mathrm{C\,m^{-2}}$, and configurations with different polarization have extensively overlapping energies.
Trajectories (see SI~\cite{supp}) show frequent switching among distinct low-polarization hydrogen-bond configurations, indicative of proton disorder.
Finally, note that in a finite supercell, $P$ takes discrete values because each winding-loop update changes the supercell dipole by a finite step under periodic boundary conditions.

\vspace{1em}
\textbf{Discussion}

In summary, our results represent a leap in \emph{ab initio} simulations of proton-ordering transitions in water ice, demonstrating that hydrogen-bond configurations under local ice-rule constraints are now tractable in large periodic supercells.
We achieve this through large-scale Monte Carlo sampling, with millions of thinned samples per temperature for systems of up to $360$ water molecules, enabled by a high-accuracy equivariant MLIP~\cite{batatia2022mace, batatia2025design}.
The model correctly identifies the ferroelectric ground state~\cite{leadbetter1985equilibrium, jackson1997single} and resolves meV-scale energetic orderings that are often reported inconsistently in previous simulations~\cite{rick2003dielectric, hirsch2004quantum, raza2011proton,piaggi2021enhancing, zhang2021phase}.
The sampling framework combines nonlocal loop updates with MALA moves, strictly enforcing the ice rules while sampling both discrete hydrogen-bond configurations and continuous atomic coordinates, as well as lattice degrees of freedom.
We find that the proton-ordering transition between ice Ih and XI is first order, with a transition temperature of $T_{\mathrm{c}} \approx 83~\mathrm{K}$.

Two sources of uncertainty remain in the $T_{\mathrm{c}}$ obtained with the current approach.
First, nuclear quantum effects are not included in the present classical Monte Carlo simulations. Experimentally, the $\mathrm{H_2O}$-to-$\mathrm{D_2O}$ isotope shift raises $T_{\mathrm{c}}$ by about $4.8~\mathrm{K}$~\cite{yen2015proton}.
A simple harmonic estimate using square-root mass scaling of vibrational frequencies gives an isotope shift of about $6~\mathrm{K}$ and suggests that NQEs reduce $T_{\mathrm{c}}$ by $10$--$20~\mathrm{K}$~\cite{pamuk2015electronic} depending on the exchange-correlation functional.
Applying the same estimate to the MACE model yields a consistent isotope shift and indicates an NQE-induced reduction of approximately $20~\mathrm{K}$.
We therefore expect that including NQEs would lower our predicted $T_{\mathrm{c}}$ toward the experimental value of $72~\mathrm{K}$~\cite{kawada1972dielectric, tajima1982phase, tajima1984calorimetric, yen2015proton}.
Extending the sampling to path-integral Monte Carlo would raise the computational cost by several orders of magnitude~\cite{fanourgakis2009fastpimc, herrero2011isotopeiceih}, which is beyond what we can afford now. 
Second, residual exchange-correlation functional errors remain another important source of uncertainty.
Different functionals yield different mass densities for water ice~\cite{rego2020elastic} and may shift the transition temperature by about $10~\mathrm{K}$~\cite{schoenherr2014dielectric}.

\vspace{1em}
{\large\bfseries Methods}

\textbf{First-principles calculations}

Electronic structure calculations are carried out within density functional theory using the meta-generalized gradient approximation (meta-GGA) regularized-restored strongly constrained and appropriately normed ($\mathrm{r^2SCAN}$) exchange-correlation functional~\cite{furness2020r2scan}.
Calculations are performed with the Vienna Ab initio Simulation Package (VASP, version 6.1)~\cite{kresse1996vasp} with projector augmented wave pseudopotentials.
A plane-wave cutoff of $800\,\mathrm{eV}$, a $k$-point spacing of $0.5~\text{\AA}^{-1}$, and an electronic convergence criterion of $10^{-6}~\mathrm{eV}$ were used to ensure convergence.
We construct a large dataset spanning cells containing $16$ to $128$ $\mathrm{H_2O}$ molecules and thermal configurations over temperatures from $10$ to $250~\mathrm{K}$, covering densities between $900$ and $1000~\mathrm{kg/m^{3}}$.
The dataset is designed to explore a broad and diverse set of hydrogen-bond configurations.
In total, the dataset comprises $44{,}879$ configurations, of which $36{,}352$ are used for training, $4{,}039$ for validation, and $4{,}488$ for testing.
Of these, $9{,}906$ configurations contain $128$ $\mathrm{H_2O}$ molecules to ensure adequate sampling of large-system behavior and to fully capture the energy variations arising from different water molecular orientations.

\vspace{1em}
\textbf{Machine learning potential}

Using the DFT-r$^2$SCAN dataset, we train an equivariant machine-learning potential based on the multi-atomic cluster expansion (MACE)~\cite{batatia2022mace, batatia2025design}.
The model employs two message-passing layers with channels $128\times0e + 128\times1o$, a correlation order of $3$, and a spherical-harmonics expansion truncated at order $3$.
A radial cutoff of $4.5\,\text{\AA}$ is used, yielding an effective receptive field of $9\,\text{\AA}$ for each atom after two message-passing layers.
The model has $6.5\times10^{5}$ parameters and is implemented in \texttt{float32} precision to accelerate inference.
The \texttt{cuEquivariance}~\cite{cuequivariance} package is used to accelerate the model evaluation.
Training uses a weighted energy-forces loss with energy weight $10$ and forces weight $100$, and takes approximately $8$ hours on a single NVIDIA RTX 5090 GPU.
For production runs, we build the neighbor list with PyTorch~\cite{paszke2019pytorch} and perform MACE inference entirely on the GPU~\cite{supp}.
On a single NVIDIA RTX 4090 GPU, a single inference call for an $N=360$ supercell takes approximately $11~\mathrm{ms}$ for energy only and $26~\mathrm{ms}$ for energy and forces, sufficient to support large-scale Monte Carlo simulations.

\vspace{1em}
\textbf{Monte Carlo protocol}

Sampling alternates between three complementary Monte Carlo updates: loop updates~\cite{rahman1972proton, rick2003dielectric}, MALA updates~\cite{roberts1996exponential}, and cell updates~\cite{frenkel2023understanding}.
Simulations use periodic supercells with $N=64$, $96$, $128$, and $360$ water molecules at temperatures spanning $20$--$200~\mathrm{K}$, with finer $2.5~\mathrm{K}$ intervals in the transition region.
For $N=64$, $96$, and $128$, each Monte Carlo block performs $20$ loop updates followed by $25$ continuous updates. 
Each continuous update is randomly selected to be either a MALA update (with probability $90\%$) or a cell update (with probability $10\%$), and one configuration is recorded after the $25$ continuous updates.
For $N=360$, the sampling is increased to $100$ loop updates and $30$ continuous updates per block to enhance exploration of hydrogen-bond configurations.
These specific sampling intervals and probabilities are chosen to optimize efficiency.
In the long-trajectory limit, the stationary distribution is independent of these choices and depends only on the target Boltzmann distribution.
On an NVIDIA RTX 5090 GPU, generating one recorded $N=360$ sample takes approximately $1.3~\mathrm{s}$ with $130$ MACE model evaluations.

\vspace{1em}
\textbf{Dipole-based order parameters}

To detect the orientational ordering transition, we define dipole-based order parameters from the total dipole moment $\bm{M}=\sum_{i=1}^{N}\bm{\mu}_i$ and the polarization magnitude
$ P=\left| {\bm{M}}/{\Omega} \right|$,
where $\Omega$ is the instantaneous volume.
After the atomic coordinates are sampled with MACE model, we estimate per-molecule dipoles $\bm{\mu}_i$ by assigning TIP4P/Ice point charges to each configuration~\cite{abascal2005tip4pice}, placing the negative charge at a fixed offset from the oxygen along the molecular bisector.
This charge-based estimate serves solely as a convenient order parameter for detecting the phase transition and is not intended for quantitative comparison with experimental absolute polarization values.
To characterize the transition order, we compute the Binder cumulant~\cite{binder1984finite,challa1986finite} from the total dipole magnitude $M=|\bm{M}|$,
\begin{equation}
B=\frac{5}{2}-\frac{3}{2}\frac{\langle M^{4}\rangle}{\langle M^{2}\rangle^{2}},
\end{equation}
where $\langle\cdot\rangle$ denotes thermal averages over Monte Carlo samples.
The normalization coefficients are adapted for a three-dimensional vector order parameter and differ from the standard scalar definition.
In the thermodynamic limit, $B=1$ in the ordered phase and $B= 0$ in the disordered phase.

\vspace{1em}
\textbf{Enthalpy and heat capacity}

In Monte Carlo sampling, hydrogen-bond configurations and atomic coordinates are updated without explicit momenta, so the kinetic-energy contribution is neglected here.
A periodic supercell containing $N$ water molecules is simulated, with each molecule treated as $3$ unconstrained classical atoms, giving $9N$ degrees of freedom.
The potential energy $E$ contains a large harmonic vibrational contribution of $\frac{9}{2}N k_{\mathrm B}T$, in addition to anharmonic and hydrogen-bond ordering contributions.
This vibrational term forms a large background against which small energy differences between proton-ordering states must be resolved.
In the isothermal--isobaric ensemble at external pressure $\mathcal{P}$, we define
$\mathcal{H}=E+\mathcal{P}\Omega$,
that is, the enthalpy without the kinetic term, where $\Omega$ is the instantaneous cell volume.
We estimate the constant-pressure heat capacity per molecule from enthalpy fluctuations~\cite{frenkel2023understanding}:
\begin{equation}
C_{\mathcal{P}}=\frac{\langle \mathcal{H}^{2}\rangle-\langle \mathcal{H}\rangle^{2}}{N k_{\mathrm B}T^{2}}.
\end{equation}
We set $\mathcal{P}=0$ for convenience, so that these fluctuations are entirely determined by the sampled potential energy.

\vspace{1em}
{\large\bfseries Data availability}

The DFT data, the MACE model, and the Monte Carlo results are available at \url{https://github.com/zhangqi94/watericeIh}.

\vspace{1em}
{\large\bfseries Code availability}

The code for this work is openly available at \url{https://github.com/zhangqi94/watericeIh}.

\vspace{1em}
{\large\bfseries Acknowledgments}

The authors acknowledge valuable discussions with Juan Carrasquilla, Lode Pollet, Han Wang, Yuan Wan, Zhendong Cao and Wenjun Tang. The DFT calculations were performed on the supercomputing system of the Huairou Materials Genome
Platform.
This work is supported by the National Natural Science Foundation of China under Grants No. T2225018, No. 12188101, No. T2121001, the Cross-Disciplinary Key Project of Beijing Natural Science Foundation No. Z250005, the Strategic Priority Research Program of the Chinese Academy of Sciences under Grants No. XDB0500000, and the National Key Projects for Research and Development of China Grants No. 2021YFA1400400.

\vspace{1em}
{\large\bfseries Competing interests}

The authors declare no competing interests.

\vspace{1em}
{\large\bfseries Additional information}

{\bf Supplementary Information}
The online version contains supplementary information.

\bibliography{article_arxiv.bbl}


\clearpage
\onecolumngrid
\begin{center}
\makeatletter
{\Large\bfseries Supplementary Information: \@title\par}
\makeatother
\end{center}
\vspace{0.5em}
\setcounter{figure}{0}
\setcounter{table}{0}
\setcounter{equation}{0}
\renewcommand{\thefigure}{S\arabic{figure}}
\renewcommand{\thetable}{S\arabic{table}}
\renewcommand{\theequation}{S\arabic{equation}}

\renewcommand{\theHfigure}{S\arabic{figure}}
\renewcommand{\theHtable}{S\arabic{table}}
\renewcommand{\theHequation}{S\arabic{equation}}
\normalfont\normalsize

\newcounter{sisec}
\renewcommand{\thesisec}{S\arabic{sisec}}
\newcommand{\SIheading}[1]{%
  \refstepcounter{sisec}%
  {\noindent\large\bfseries \thesisec\quad #1\par}%
  \normalfont\normalsize
  \vspace{0.5em}%
}

\vspace{1.0em}

\SIheading{Previous studies and challenges}

The ice Ih to XI transition has been studied experimentally for several decades.
KOH doping introduces defects and promotes proton rearrangement in water ice, reducing the proton ordering time from years to days.
Adiabatic calorimetry measurements on KOH-doped samples revealed a heat capacity anomaly near $72$~K, indicating a first-order transition~\cite{tajima1982phase}.
Dielectric constant measurements on pure samples showed different transition temperatures upon heating and cooling, with the Ih to XI transition beginning at $73.4$~K and the XI to Ih transition at $58.9$~K~\cite{yen2015proton}.
The transition temperature is essentially independent of pressure.
Moreover, isotope experiments found that $\mathrm{D_2O}$ ice exhibits a transition temperature $4.8$~K higher than $\mathrm{H_2O}$ ice~\cite{yen2015proton}.
Neutron diffraction experiments determined the crystal structure of the low-temperature ordered phase, identifying ice XI as a ferroelectric $Cmc2_1$ structure~\cite{leadbetter1985equilibrium, howe1989adetermination, jackson1997single}.

Computational studies of the Ih to XI transition face several challenges.
First, the potential energy surface (PES) must achieve meV-per-$\mathrm{H_2O}$ accuracy to correctly rank competing hydrogen-bond configurations and identify the ferroelectric $Cmc2_1$ structure as the ground state, as verified by both density functional theory (DFT) calculations~\cite{hirsch2004quantum, raza2011proton} and experiments~\cite{leadbetter1985equilibrium, howe1989adetermination, jackson1997single}.
Yet energetic accuracy alone is not sufficient.
Second, the exponentially large ice-rule-constrained configuration space must be sampled thoroughly, yet eV-scale barriers between distinct hydrogen-bond topologies create severe kinetic bottlenecks that trap conventional molecular dynamics in single orderings.
At the same time, simultaneously incorporating hydrogen-bond configuration sampling and lattice vibrations remains difficult, and previous approaches differ in how they address these challenges.
Previous computational attempts~\cite{tajima1982phase,barkema1993properties,rick2003dielectric,singer2005hbtopology,knight2006hbtopology,schoenherr2014dielectric,piaggi2021enhancing,zhang2021phase} are summarized in Table~\ref{tab:previous_studies}.

\begin{table}[htbp]
\caption{\textbf{Summary of previous studies of the ice Ih to XI transition.}
``Ground state structure" indicates whether the potential energy surface stabilizes the ferroelectric $Cmc2_1$ structure.
``Ice-rule configuration" indicates whether hydrogen-bond configurations that obey the ice rule are treated by explicit equilibrium sampling.
``Atomic vibration" indicates whether atomic displacements from lattice vibrations are included.
``System size" reports the number of water molecules in the simulation cell.
}
\centering
\small
\begin{tabular}{lllllll}
\hline
\hline
\noalign{\vskip 2pt}
\parbox[c]{3.2cm}{\centering Reference} &
\parbox[c]{3.3cm}{\centering Potential energy surface (model)} &
\parbox[c]{2.5cm}{\centering Ground state structure} &
\parbox[c]{2.5cm}{\centering Ice-rule configuration} &
\parbox[c]{1.9cm}{\centering Atomic vibration} &
\parbox[c]{1.2cm}{\centering System size} &
\parbox[c]{2.2cm}{\centering Transition temperature} \\
\noalign{\vskip 2pt}
\hline
\noalign{\vskip 2pt}
\parbox[c]{3.2cm}{\centering Tajima (1982)~\cite{tajima1982phase}} &
\parbox[c]{3.3cm}{\centering Experiment} &
\parbox[c]{2.5cm}{\centering $Cmc2_1$} &
\parbox[c]{2.5cm}{\centering -} &
\parbox[c]{1.9cm}{\centering -} &
\parbox[c]{1.2cm}{\centering $\infty$} &
\parbox[c]{2.2cm}{\centering 72 K} \\
\hline
\noalign{\vskip 2pt}
\parbox[c]{3.2cm}{\centering Barkema (1993)~\cite{barkema1993properties}} &
\parbox[c]{3.3cm}{\centering Point charge} &
\parbox[c]{2.5cm}{\centering \xmark} &
\parbox[c]{2.5cm}{\centering \cmark} &
\parbox[c]{1.9cm}{\centering \xmark} &
\parbox[c]{1.2cm}{\centering 1400} &
\parbox[c]{2.2cm}{\centering 36 K} \\
\parbox[c]{3.2cm}{\centering Rick (2003)~\cite{rick2003dielectric}} &
\parbox[c]{3.3cm}{\centering TIP4P} &
\parbox[c]{2.5cm}{\centering \xmark} &
\parbox[c]{2.5cm}{\centering \cmark} &
\parbox[c]{1.9cm}{\centering \cmark} &
\parbox[c]{1.2cm}{\centering 360} &
\parbox[c]{2.2cm}{\centering 10--50 K} \\
\parbox[c]{3.2cm}{\centering Rick (2003)~\cite{rick2003dielectric}} &
\parbox[c]{3.3cm}{\centering TIP4P-FQ} &
\parbox[c]{2.5cm}{\centering \xmark} &
\parbox[c]{2.5cm}{\centering \cmark} &
\parbox[c]{1.9cm}{\centering \cmark} &
\parbox[c]{1.2cm}{\centering 360} &
\parbox[c]{2.2cm}{\centering 50--100 K} \\
\parbox[c]{3.2cm}{\centering Singer (2005)~\cite{singer2005hbtopology}} &
\parbox[c]{3.3cm}{\centering BLYP (Graph invariant)} &
\parbox[c]{2.5cm}{\centering \cmark} &
\parbox[c]{2.5cm}{\centering \cmark} &
\parbox[c]{1.9cm}{\centering \xmark} &
\parbox[c]{1.2cm}{\centering 896} &
\parbox[c]{2.2cm}{\centering 98 K} \\
\parbox[c]{3.2cm}{\centering Sch\"{o}nherr (2014)~\cite{schoenherr2014dielectric}} &
\parbox[c]{3.3cm}{\centering PBE-D2 (DFT)} &
\parbox[c]{2.5cm}{\centering \cmark} &
\parbox[c]{2.5cm}{\centering \cmark} &
\parbox[c]{1.9cm}{\centering \cmark} &
\parbox[c]{1.2cm}{\centering 96} &
\parbox[c]{2.2cm}{\centering 90--100 K} \\
\parbox[c]{3.2cm}{\centering Sch\"{o}nherr (2014)~\cite{schoenherr2014dielectric}} &
\parbox[c]{3.3cm}{\centering PBE0-D2 (DFT)} &
\parbox[c]{2.5cm}{\centering \cmark} &
\parbox[c]{2.5cm}{\centering \cmark} &
\parbox[c]{1.9cm}{\centering \cmark} &
\parbox[c]{1.2cm}{\centering 96} &
\parbox[c]{2.2cm}{\centering 70--80 K} \\
\parbox[c]{3.2cm}{\centering Zhang (2021)~\cite{zhang2021phase}} &
\parbox[c]{3.3cm}{\centering SCAN (Deep Potential)} &
\parbox[c]{2.5cm}{\centering \xmark} &
\parbox[c]{2.5cm}{\centering \xmark} &
\parbox[c]{1.9cm}{\centering \cmark} &
\parbox[c]{1.2cm}{\centering 128} &
\parbox[c]{2.2cm}{\centering 66 K} \\
\parbox[c]{3.2cm}{\centering \textbf{This work}} &
\parbox[c]{3.3cm}{\centering \textbf{r$^{2}$SCAN (MACE)}} &
\parbox[c]{2.5cm}{\centering \cmark} &
\parbox[c]{2.5cm}{\centering \cmark} &
\parbox[c]{1.9cm}{\centering \cmark} &
\parbox[c]{1.2cm}{\centering \textbf{360}} &
\parbox[c]{2.2cm}{\centering \textbf{83 K}} \\
\noalign{\vskip 2pt}
\hline
\hline
\end{tabular}
\label{tab:previous_studies}
\end{table}

Empirical water potentials such as the transferable intermolecular potential with 4 points (TIP4P) and TIP4P with fluctuating charges (TIP4P-FQ)~\cite{jorgensen1983tip4p, rick1994tip4pfq} are computationally inexpensive.
TIP4P is a nonpolarizable fixed-charge model, whereas TIP4P-FQ incorporates polarization through fluctuating charges, but both produce energetic orderings of hydrogen-bond configurations that are inconsistent with density functional theory (DFT)~\cite{hirsch2004quantum}.
In particular, neither model can capture the meV-scale energy differences between competing proton orderings due to limitations in their parameterization and treatment of many-body interactions.
They also fail to reproduce the ferroelectric $Cmc2_1$ structure as the ground state, instead favoring antiferroelectric order~\cite{rick2003dielectric}.
Predicted transition temperatures vary widely across different empirical potentials, reflecting their inability to correctly describe the subtle energetics of proton ordering.

Graph-invariant approaches~\cite{singer2005hbtopology,knight2006hbtopology} fit a model that depends on the hydrogen-bond configuration by calibrating it against DFT Becke-Lee-Yang-Parr (BLYP)~\cite{becke1988density,lee1988development} energies, improving the energetic ordering of distinct topologies.
However, the model depends only on topology and maps each hydrogen-bond configuration to a single energy, so different atomic configurations within the same topology are not distinguished and vibrational contributions are neglected.
The authors acknowledge this limitation, noting that geometric relaxation introduces systematic energy differences, but assume that the vibrational energies are similar across isomers and therefore omit them from the calculation.
Because vibrations broaden the energy distributions of different hydrogen-bond topologies and can cause them to overlap, this approximation may introduce uncontrolled errors in the predicted transition temperature.

Monte Carlo sampling with on-the-fly DFT energy evaluations~\cite{schoenherr2014dielectric} is computationally expensive to access larger system sizes and finer temperature resolution, limiting the simulations to $96$ water molecules and $10$~K temperature intervals in the transition region.
To reduce computational cost, the method uses a composite approach where an entire sub-Markov chain is generated using empirical force fields, and only a single DFT calculation is needed to accept or reject the entire chain.
The study used the dielectric constant to characterize the transition, but finite-size limitations prevent observation of the divergence expected in the thermodynamic limit, and the authors note that size effects cannot be excluded near the transition.
Moreover, they found that the Perdew-Burke-Ernzerhof with Grimme's second-generation dispersion correction (PBE-D2)~\cite{perdew1996generalized, grimme2006semiempirical} functional yields a transition temperature about $20$~K higher than the hybrid functional PBE0-D2~\cite{adamo1999toward, grimme2006semiempirical}.

Machine-learning interatomic potentials (MLIPs) offer near-DFT accuracy at much lower computational cost.
In Ref.~\cite{piaggi2021enhancing}, enhanced sampling with a bias potential was used to promote ionic defect formation, accelerating proton rearrangement in molecular dynamics with a machine learning potential originally developed for liquid water studies.
However, the Deep Potential model was not optimized for ice hydrogen-bond ordering, and the training set did not include proton transfer configurations, resulting in a PES that is not a perfect representation of the DFT energies.
The study did not directly simulate the phase transition, instead inferring the transition temperature by extrapolation from high-temperature simulations.
Due to the lack of data below $200$~K, they rely on extrapolation from $200$--$300$~K, yielding a rough estimate of the transition temperature around $50$~K, or $67.5$~K based on $0$~K enthalpy differences.

In Ref.~\cite{zhang2021phase}, another Deep Potential model was trained to study the phase diagram of water, but according to our own tests, the model does not stabilize the $Cmc2_1$ ground state identified by DFT and experiments.
This is likely because the study focuses on constructing a comprehensive phase diagram of water over a broad range of temperatures and pressures, rather than specifically investigating the proton-ordering transition.
For the Ih to XI transition, the free-energy comparison treats the configurational entropy of disordered Ih by directly adding the Pauling residual entropy~\cite{pauling1935structure, macdowell2004combinatorial, herrero2013configurational, berg2007residual, xu2025equivalence}.
As a result, it does not explicitly sample the configurational space and neglects possible variations in the residual entropy and degree of disorder with thermodynamic state.
The authors also note that the model accuracy may not be sufficient for resolving the meV-scale energy differences required for proton-ordering simulations.

\vspace{1.0em}
\SIheading{Monte Carlo sampling algorithm}

Evaluations of the MACE potential dominate the computational cost, accounting for over $90\%$ of the runtime.
We implement the Monte Carlo driver in \texttt{NumPy}~\cite{numpy} to handle proposal generation, acceptance decisions, and data recording with minimal additional overhead.
As noted in the main text, ergodicity is ensured by alternating three complementary moves that sample both discrete and continuous degrees of freedom.
Metropolis short-loop moves~\cite{rahman1972proton, rick2003dielectric} rearrange the discrete hydrogen-bond topology.
MALA moves~\cite{roberts1996exponential} and cell-length updates~\cite{frenkel2023understanding} sample the continuous degrees of freedom, thermally exploring atomic coordinates and cell lengths at the target temperature and pressure. 
Each proposed move is accepted or rejected to enforce the target Boltzmann distribution.
The hydrogen-bond topology is encoded by a binary state vector $\bm{S}\in\{0,1\}^{\otimes 2N}$, where $N$ is the number of water molecules in the supercell.
For each nearest-neighbor oxygen pair $(i,j)$, we set $S_{ij}=0$ if the proton lies closer to $i$ and $S_{ij}=1$ if it lies closer to $j$.
We denote the atomic coordinates by $\bm{X}$ and use an orthorhombic simulation cell with side lengths $\bm{L}$.
The potential energy surface is defined using the MACE model, from which we compute the potential energy $E=V_{\mathrm{MACE}}(\bm{X},\bm{L})$ and forces $\bm{F}=-\nabla_{\bm{X}}V_{\mathrm{MACE}}(\bm{X},\bm{L})$.

\begin{algorithm}[ht]
\caption{Metropolis short-loop updates of the hydrogen-bond configuration}
\label{alg:loop}
\raggedright
\textbf{Inputs:} Hydrogen-bond state $\bm{S}$, atomic coordinates $\bm{X}$, cell lengths $\bm{L}$, temperature $T$. \hfill $\triangleright\ \bm{S}\in\{0,1\}^{\otimes 2N},\ \bm{X}\in\mathbb{R}^{3N\times 3},\ \bm{L}\in\mathbb{R}^{3}$\\
\textbf{Outputs:} Hydrogen-bond state $\bm{S}$, atomic coordinates $\bm{X}$, potential energy $E$.
\begin{algorithmic}[1]
\STATE Set the inverse temperature $\beta=(k_{\mathrm{B}}T)^{-1}$.
\STATE Compute the initial potential energy $E=V_{\mathrm{MACE}}(\bm{X},\bm{L})$. \hfill $\triangleright\ E\in\mathbb{R}$
  \STATE Randomly select a nearest-neighbor oxygen pair $(i,j)$ and initialize the walk record $\mathcal{P}=[i,j]$.
  \REPEAT
      \IF{$S_{ij}=0$}
        \STATE Randomly choose $k$ from the two neighbors of $j$ if the hydrogen on bond $j$--$k$ is closer to $j$.
      \ELSIF{$S_{ij}=1$}
        \STATE Randomly choose $k$ from the two neighbors of $j$ if the hydrogen on bond $j$--$k$ is closer to $k$.
      \ENDIF
    \STATE Set $i\leftarrow j$ and $j\leftarrow k$.
    \STATE Append $j$ to the walk record $\mathcal{P}$.
	  \UNTIL{$j$ revisits an earlier site in $\mathcal{P}$}
		\STATE Identify the closed loop $\mathcal{L}$ as the segment of $\mathcal{P}$ between the first and second visits to site $j$.
	  \STATE Flip the hydrogens along $\mathcal{L}$ to obtain $(\bm{S}',\bm{X}')$ and compute the trial potential energy $E'=V_{\mathrm{MACE}}(\bm{X}',\bm{L})$. \hfill $\triangleright\ E'\in\mathbb{R}$
	  \STATE Draw $u\sim \mathcal{U}(0,1)$ uniformly at random. \hfill $\triangleright\ u\in(0,1)$
	  \IF{$\ln u < -\beta(E'-E)$}
	    \STATE Accept the move and set $(\bm{S},\bm{X},E)\leftarrow(\bm{S}',\bm{X}',E')$.
	  \ENDIF
\end{algorithmic}
\end{algorithm}

The short-loop update of the hydrogen-bond topology is summarized in Algorithm~\ref{alg:loop} and is denoted by \texttt{loop\_move}~\cite{rahman1972proton, rick2003dielectric}.
Starting from a randomly chosen nearest-neighbor oxygen pair $(i,j)$, we generate a directed random walk on the hydrogen-bond network.
At each step, the next oxygen $k$ is drawn uniformly from the two candidates consistent with the current orientation $S_{ij}$.
Specifically, the chosen neighbor is such that the hydrogen on bond $j$-$k$ is closer to $j$ when $S_{ij}=0$ and closer to $k$ when $S_{ij}=1$.
The walk terminates upon revisiting an oxygen, which defines a closed loop $\mathcal{L}$.
We flip the hydrogens along $\mathcal{L}$ to propose a move to $(\bm{S}',\bm{X}')$, which is then accepted or rejected according to the Metropolis criterion.
The loop update preserves the ice rules and enables ergodic sampling within the ice-rule manifold.
Moreover, updating the atomic coordinates during a loop flip requires care.
A mirror reflection of hydrogen positions across an O-O bond breaks the intramolecular geometry.
This distorts the H-O-H angles by tens of degrees, raises the energy by hundreds of meV, and drives the acceptance rate nearly to zero.
Following Ref.~\cite{rick2003dielectric}, we instead rotate each water molecule on $\mathcal{L}$ as a rigid body about its dipole axis.
The rotation angle is determined from the relative orientations of neighboring molecules before and after the flip, so that for each water molecule, the O-H bond lengths and H-O-H angle remain unchanged.
This preserves intramolecular geometry while allowing hydrogen-bond rearrangements.

\begin{algorithm}[ht]
\caption{Metropolis-adjusted Langevin update of atomic coordinates}
\label{alg:mala}
\raggedright
\textbf{Inputs:} Atomic coordinates $\bm{X}$, cell lengths $\bm{L}$, temperature $T$, step width $\sigma_H$. \hfill $\triangleright\ \bm{X}\in\mathbb{R}^{3N\times 3},\ \bm{L}\in\mathbb{R}^{3}$\\
\textbf{Outputs:} Atomic coordinates $\bm{X}$, potential energy $E$, forces $\bm{F}$.
\begin{algorithmic}[1]
\STATE Set the inverse temperature $\beta=(k_{\mathrm{B}}T)^{-1}$.
\STATE Set the per-atom step widths $\sigma_i=\sigma_H$ for hydrogens and $\sigma_i=0.25\sigma_H$ for oxygens.
\STATE Compute the potential energy $E=V_{\mathrm{MACE}}(\bm{X},\bm{L})$ and forces $\bm{F}=-\nabla_{\bm{X}}V_{\mathrm{MACE}}(\bm{X},\bm{L})$. \hfill $\triangleright\ E\in\mathbb{R},\ \bm{F}\in\mathbb{R}^{3N\times 3}$
  \STATE Draw independent standard Gaussian noise vectors $\bm{\eta}_i \sim \mathcal{N}(\bm{0}, \bm{I}_3)$ for all atoms $i$. \hfill $\triangleright\ \bm{\eta}\in\mathbb{R}^{3N\times 3}$
  \STATE Propose $\bm{X}'_i=\texttt{wrap}\!\left(\bm{X}_i+\tfrac{1}{2}\sigma_i^2\,\beta\,\bm{F}_i+\sigma_i\bm{\eta}_i\right)$, wrapping the coordinates back into the cell. \hfill $\triangleright\ \bm{X}'\in\mathbb{R}^{3N\times 3}$
  \STATE Compute the trial potential energy $E'=V_{\mathrm{MACE}}(\bm{X}',\bm{L})$ and forces $\bm{F}'=-\nabla_{\bm{X}'}V_{\mathrm{MACE}}(\bm{X}',\bm{L})$. \hfill $\triangleright\ E'\in\mathbb{R},\ \bm{F}'\in\mathbb{R}^{3N\times 3}$
  \STATE Compute the minimum-image displacement $\Delta\bm{X}=\texttt{mic}(\bm{X}'-\bm{X})$. \hfill $\triangleright\ \Delta\bm{X}\in\mathbb{R}^{3N\times 3}$
  \STATE Compute the forward log proposal density $\ln q_{\mathrm{fwd}}=-\tfrac{1}{2}\sum_{i,\alpha}\!\left[\frac{\!\left(\Delta X_{i\alpha}-\tfrac{1}{2}\sigma_i^2\beta F_{i\alpha}\right)^{2}}{\sigma_i^{2}}+\ln(2\pi\sigma_i^{2})\right]$. \hfill $\triangleright\ \ln q_{\mathrm{fwd}}\in\mathbb{R}$
  \STATE Compute the backward log proposal density $\ln q_{\mathrm{bwd}}=-\tfrac{1}{2}\sum_{i,\alpha}\!\left[\frac{\!\left(-\Delta X_{i\alpha}-\tfrac{1}{2}\sigma_i^2\beta F'_{i\alpha}\right)^{2}}{\sigma_i^{2}}+\ln(2\pi\sigma_i^{2})\right]$. \hfill $\triangleright\ \ln q_{\mathrm{bwd}}\in\mathbb{R}$
  \STATE Draw $u\sim\mathcal{U}(0,1)$ uniformly at random. \hfill $\triangleright\ u\in(0,1)$
  \IF{$\ln u < (\ln q_{\mathrm{bwd}}-\ln q_{\mathrm{fwd}})-\beta(E'-E)$}
    \STATE Accept the move and set $(\bm{X},E,\bm{F})\leftarrow(\bm{X}',E',\bm{F}')$.
  \ENDIF
\end{algorithmic}
\end{algorithm}

Algorithm~\ref{alg:mala} summarizes the Metropolis-adjusted Langevin algorithm (MALA) used for atomic coordinate updates, which we denote by \texttt{mala\_move}~\cite{roberts1996exponential}.
The MALA proposal combines a drift term proportional to the force with Gaussian noise, biasing moves toward lower-energy configurations and accelerating exploration relative to pure diffusion.
Because the drift depends on the current force, the proposal distribution is asymmetric and therefore requires a Metropolis--Hastings correction based on the forward and backward proposal densities.
We move all atoms using species-dependent step widths chosen to scale approximately as $\sigma \propto m^{-1/2}$, with $\sigma_O=0.25\sigma_H$.
The ratio is chosen because the much larger oxygen mass leads to a smaller trial displacement, consistent with the physical expectation that lighter atoms undergo larger thermal displacements.
This scaling ensures consistent sampling of the thermal distributions of both oxygen and hydrogen while avoiding overly large trial displacements for the heavier oxygen atoms.
We further set $\sigma_H\propto\sqrt{T}$ to match the temperature dependence of thermal motion.

\begin{algorithm}[ht]
\caption{Cell lengths update via log-space proposals with Metropolis--Hastings acceptance}
\label{alg:cell}
\raggedright
\textbf{Inputs:} Atomic coordinates $\bm{X}$, cell lengths $\bm{L}$, temperature $T$, pressure $\mathcal{P}$, step width $\sigma_{\mathrm{cell}}$. \hfill $\triangleright\ \bm{X}\in\mathbb{R}^{3N\times 3},\ \bm{L}\in\mathbb{R}^{3}$\\
\textbf{Outputs:} Atomic coordinates $\bm{X}$, cell lengths $\bm{L}$, potential energy $E$.
\begin{algorithmic}[1]
\STATE Set the inverse temperature $\beta=(k_{\mathrm{B}}T)^{-1}$.
\STATE Compute the initial potential energy $E=V_{\mathrm{MACE}}(\bm{X},\bm{L})$ and cell volume $\Omega = L_a L_b L_c$. \hfill $\triangleright\ E\in\mathbb{R},\ \Omega\in\mathbb{R}$
  \STATE Draw a standard Gaussian noise vector $\bm{\xi}\sim\mathcal{N}(\bm{0},\bm{I}_3)$. \hfill $\triangleright\ \bm{\xi}\in\mathbb{R}^{3}$
  \STATE Propose $L'_{\alpha}=L_{\alpha}\exp(\sigma_{\mathrm{cell}}\xi_{\alpha})$ for $\alpha\in\{a,b,c\}$ and scale coordinates $X'_{i\alpha}=X_{i\alpha}L'_\alpha/L_\alpha$ for all atoms $i$. \hfill $\triangleright\ \bm{L}'\in\mathbb{R}^{3},\ \bm{X}'\in\mathbb{R}^{3N\times 3}$
  \STATE Compute potential energy $E'=V_{\mathrm{MACE}}(\bm{X}',\bm{L}')$ and proposed cell volume $\Omega' = L_a' L_b' L_c'$. \hfill $\triangleright\ E'\in\mathbb{R},\ \Omega'\in\mathbb{R}$
  \STATE Draw $u\sim\mathcal{U}(0,1)$ uniformly at random. \hfill $\triangleright\ u\in(0,1)$
  \IF{$\ln u < -\beta\left[E'-E+\mathcal{P}(\Omega'-\Omega)-(3N+1)k_{\mathrm{B}}T\ln(\Omega'/\Omega)\right]$}
    \STATE Accept the move and set $(\bm{X},\bm{L},E)\leftarrow(\bm{X}',\bm{L}',E')$.
  \ENDIF
\end{algorithmic}
\end{algorithm}

To sample the isothermal-isobaric ensemble, Algorithm~\ref{alg:cell} updates the orthorhombic cell lengths $\bm{L}=(L_a,L_b,L_c)$ using a Metropolis-Hastings move, which we denote by \texttt{cell\_move}~\cite{frenkel2023understanding}.
Because the simulation cell is constrained to remain orthorhombic, only the three cell lengths are varied and no shear degrees of freedom are sampled.
We propose log-space updates of the form $L'_{\alpha}=L_{\alpha}\exp(\sigma_{\mathrm{cell}}\xi_{\alpha})$, where $\xi_{\alpha}$ is drawn from a standard Gaussian distribution, and rescale the atomic coordinates affinely so that their fractional coordinates remain unchanged.
The corresponding Metropolis-Hastings acceptance probability depends on the enthalpy change and on a Jacobian factor associated with the volume scaling, whose exponent $(3N+1)$ has two contributions.
The $3N$ term comes from the affine scaling of the $3N$ atomic coordinate components, while the additional $+1$ arises from the Jacobian of the log-space volume transformation~\cite{frenkel2023understanding}.
As in the MALA coordinate updates, we choose $\sigma_{\mathrm{cell}}\propto\sqrt{T}/\Omega^{1/3}$ to reflect the thermal scale of cell fluctuations while maintaining comparable proposal sizes across different system sizes.

\begin{algorithm}[ht]
\caption{Composite Monte Carlo sampling algorithm}
\label{alg:hybrid_mc}
\raggedright
\textbf{Inputs:} Hydrogen-bond state $\bm{S}$, atomic coordinates $\bm{X}$, cell lengths $\bm{L}$, temperature $T$, pressure $\mathcal{P}$, update counts $N_{\mathrm{discrete}}$ and $N_{\mathrm{continuous}}$, probability $p_{\mathrm{mala}}$ for MALA updates, number of sampling cycles $N_{\mathrm{cycle}}$. \hfill $\triangleright\ \bm{S}\in\{0,1\}^{\otimes 2N},\ \bm{X}\in\mathbb{R}^{3N\times 3},\ \bm{L}\in\mathbb{R}^{3}$\\
\textbf{Outputs:} Sampled trajectory of $(\bm{S},\bm{X},\bm{L})$ and observables.
\begin{algorithmic}[1]
\FOR{$c = 1$ to $N_{\mathrm{cycle}}$}
  \FOR{$m = 1$ to $N_{\mathrm{discrete}}$}
    \STATE $(\bm{S},\bm{X},E)\leftarrow$ \texttt{loop\_move}$(\bm{S},\bm{X},\bm{L},T)$ (Algorithm~\ref{alg:loop}).
  \ENDFOR
  \FOR{$n = 1$ to $N_{\mathrm{continuous}}$}
    \STATE Draw a uniform random number $w \sim \mathcal{U}(0,1)$. \hfill $\triangleright\ w\in(0,1)$
    \IF{$w < p_{\mathrm{mala}}$}
      \STATE $(\bm{X},E)\leftarrow$ \texttt{mala\_move}$(\bm{X},\bm{L},T)$ (Algorithm~\ref{alg:mala}).
    \ELSE
      \STATE $(\bm{X},\bm{L},E)\leftarrow$ \texttt{cell\_move}$(\bm{X},\bm{L},T,\mathcal{P})$ (Algorithm~\ref{alg:cell}).
    \ENDIF
  \ENDFOR
  \STATE Record sample $(\bm{S},\bm{X},\bm{L})$ and evaluate observables (energy, dipole).
\ENDFOR
\end{algorithmic}
\end{algorithm}

Algorithm~\ref{alg:hybrid_mc} summarizes the composite sampling protocol at fixed temperature $T$ and pressure $\mathcal{P}$.
Each sampling cycle consists of $N_{\mathrm{discrete}}$ loop moves that update the discrete hydrogen-bond topology, followed by $N_{\mathrm{continuous}}$ moves that sample the continuous degrees of freedom.
For each continuous move, the algorithm chooses either a MALA coordinate update or a cell-length update with probabilities $p_{\mathrm{mala}}$ and $1-p_{\mathrm{mala}}$, respectively.
After the continuous updates, one configuration is recorded and the associated observables, including the potential energy, dipole moment, and cell volume, are evaluated.
Repeating this cycle $N_{\mathrm{cycle}}$ times generates a recorded trajectory.
The initial thermalization segment is discarded, and only post-thermalization samples are retained for analysis.
To increase throughput, multiple independent Markov chains are run in parallel with independent random seeds, and the resulting samples are aggregated across chains.
Recording after the continuous updates improves sampling efficiency because the loop acceptance rate is typically below $2\%$, even at high temperature, far lower than the $40\%$--$70\%$ acceptance rates of the MALA and cell updates.
This choice avoids storing repeated configurations from rejected loop proposals and ensures that each recorded sample includes updated coordinates and cell parameters.
In the long-trajectory limit, the stationary distribution is independent of the specific values of $N_{\mathrm{discrete}}$, $N_{\mathrm{continuous}}$, and $p_{\mathrm{mala}}$, and these parameters affect only sampling efficiency and not the sampled ensemble.

\vspace{1.0em}

\SIheading{Performance optimization of graph construction}

In the multi-atomic cluster expansion (MACE) framework~\cite{batatia2022mace,batatia2025design}, each energy and force evaluation requires the construction of an atomistic graph in which edges connect pairs of atoms within a cutoff radius $r_{\mathrm{max}}$.
This input graph is specified by the neighbor-pair indices \texttt{edge\_index} together with the corresponding periodic-image translation vectors \texttt{shifts}.
In the original MACE implementation, this graph is constructed on the central processing unit (CPU) using the neighbor-list routine provided by \texttt{matscipy}~\cite{matscipy}.
In our Monte Carlo simulations, non-local loop moves simultaneously flip multiple hydrogen atoms and change neighbor relationships across many atoms, while coordinate updates modify atomic positions. 
As a result, the neighbor graph must be rebuilt at every evaluation, rather than being reused across steps as is typically done in conventional molecular dynamics simulations.
For large supercells, CPU-side graph construction can be several times more time-consuming than graphics processing unit (GPU) inference and can become the dominant bottleneck.
To eliminate this overhead, we implement a neighbor-list generator in PyTorch~\cite{paszke2019pytorch} that runs entirely on the GPU.
Following the logic of the \texttt{eval\_configs.py} script in the MACE code repository, we developed a streamlined inference function that integrates GPU-based neighbor-list construction.

\begin{algorithm}[ht]
\caption{Vectorized neighbor-list construction under periodic boundary conditions}
\label{alg:neighborlist_gpu}
\raggedright
\textbf{Inputs:} Atomic Cartesian positions $\bm{X}=\{\bm{x}_i\}$, cell matrix $\bm{A}$, cutoff $r_{\mathrm{max}}$. \hfill $\triangleright\ \bm{X}\in\mathbb{R}^{N_{\mathrm{atom}}\times 3},\ \bm{A}\in\mathbb{R}^{3\times 3}$ \\
\textbf{Outputs:} Edge list \texttt{edge\_index}, translation vectors \texttt{shifts}.
\begin{algorithmic}[1]
\STATE Construct all pairwise displacement vectors $\bm{r}_{ij}=\bm{x}_j-\bm{x}_i$ through broadcasting. \hfill $\triangleright\ \bm{r}_{ij}\in\mathbb{R}^{N_{\mathrm{atom}}\times N_{\mathrm{atom}}\times 3}$
\STATE Transform to fractional coordinates $\bm{u}_{ij}=\bm{r}_{ij}\bm{A}^{-1}$ through vectorized matrix multiplication. \hfill $\triangleright\ \bm{u}_{ij}\in\mathbb{R}^{N_{\mathrm{atom}}\times N_{\mathrm{atom}}\times 3}$
\STATE Apply the minimum-image convention: $\bm{n}_{ij}=\texttt{round}(\bm{u}_{ij})$ and $\bm{u}^{\mathrm{MIC}}_{ij}=\bm{u}_{ij}-\bm{n}_{ij}$. \hfill $\triangleright\ \bm{n}_{ij}\in\mathbb{Z}^{N_{\mathrm{atom}}\times N_{\mathrm{atom}}\times 3}$
\STATE Transform back to Cartesian coordinates $\bm{r}^{\mathrm{MIC}}_{ij}=\bm{u}^{\mathrm{MIC}}_{ij}\bm{A}$ and compute $d_{ij}^{2}=\|\bm{r}^{\mathrm{MIC}}_{ij}\|^{2}$ through vectorized operations. \hfill $\triangleright\ d_{ij}^{2}\in\mathbb{R}^{N_{\mathrm{atom}}\times N_{\mathrm{atom}}}$
\STATE Select all pairs $(i,j)$ with $d_{ij}^{2}<r_{\mathrm{max}}^{2}$ and $i\neq j$ using a boolean mask. \hfill $\triangleright\ N_{\mathrm{edge}}\ \text{pairs}$
\STATE Assemble all selected index pairs into \texttt{edge\_index}$=\{(i,j)\}$. \hfill $\triangleright\ \texttt{edge\_index}\in\mathbb{N}^{N_{\mathrm{edge}} \times 2}$
\STATE Compute the translation vectors \texttt{shifts}$_{(i,j)}=-\bm{n}_{ij}\bm{A}$ for all selected edges. \hfill $\triangleright\ \texttt{shifts}\in\mathbb{R}^{N_{\mathrm{edge}}\times 3}$
\end{algorithmic}
\end{algorithm}

Algorithm~\ref{alg:neighborlist_gpu} describes our GPU-based neighbor-list construction used to build the MACE input graph.
We denote the number of atoms by $N_{\mathrm{atom}}$ and the number of edges by $N_{\mathrm{edge}}$.
Edges are determined by the cutoff $r_{\mathrm{max}}$, so $N_{\mathrm{edge}}$ varies with the instantaneous atomic coordinates $\bm{X}$.
We follow the logic of \texttt{matscipy}~\cite{matscipy} while implementing the key steps with vectorized tensor operations to exploit GPU parallelism.
Specifically, the implementation uses broadcasting for pairwise distance computation, vectorized matrix multiplication for coordinate transformations, and masked operations for edge selection.
This eliminates CPU-side graph construction and reduces CPU-GPU data transfer and synchronization during inference.
We benchmark the implementation on a supercell containing $360$ water molecules, using an Intel i9-14900K CPU and an NVIDIA RTX 4090 GPU.
GPU-based neighbor-list construction reduces the average time for a single potential-energy evaluation from $75$~ms to $11$~ms, corresponding to a $7\times$ speedup.
The implementation is also applicable to any system under periodic boundary conditions.

\vspace{1.0em}
\SIheading{Machine learning potential benchmark}

\begin{figure*}[ht]
\centering
\includegraphics[width=0.9\columnwidth]{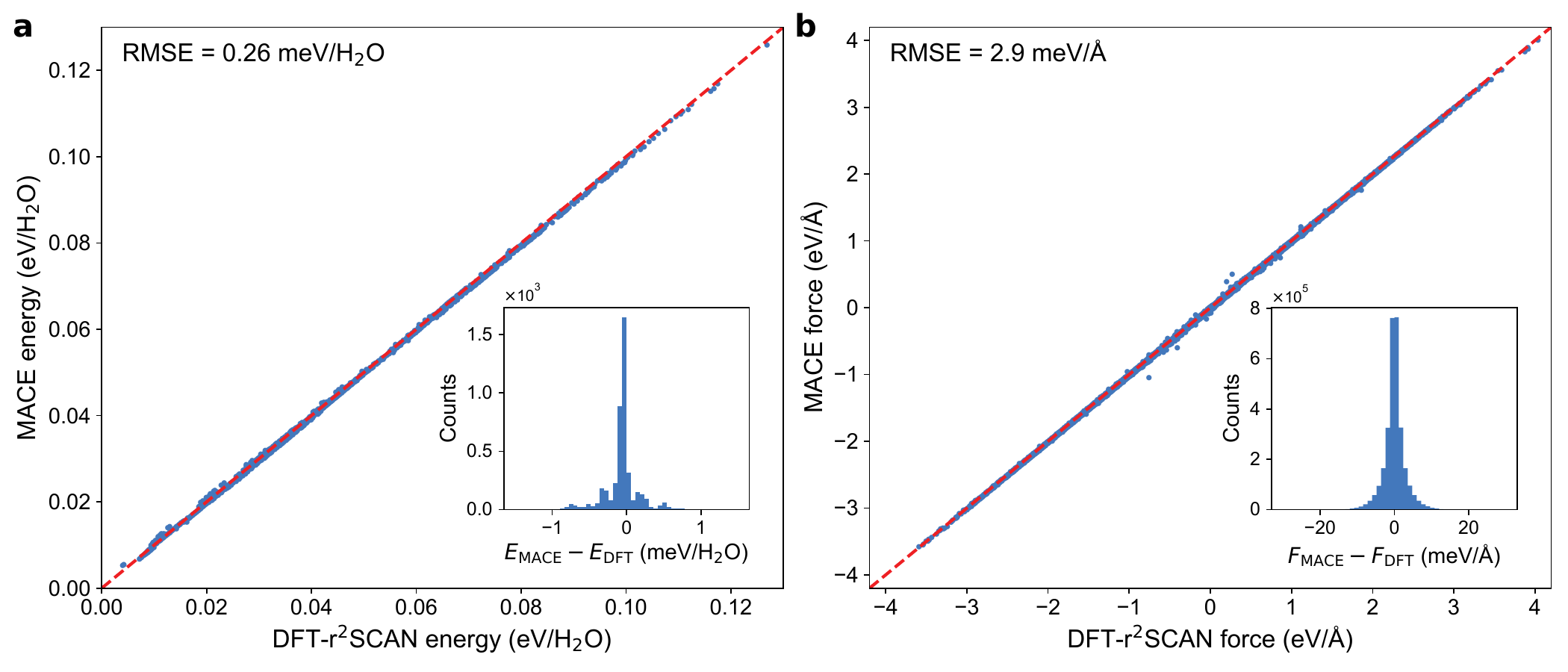}
\caption{
\textbf{Comparison between the MACE model and DFT-r$^{2}$SCAN.}
Results on the test set for (\textbf{a}) energies per water molecule and (\textbf{b}) forces.
Dots denote individual data points in the test set, and the dashed red line marks perfect agreement.
Insets show the distributions of prediction errors.
}
\label{fig:mlpes}
\end{figure*}

As shown in Fig.~\ref{fig:mlpes}, the MACE model achieves excellent agreement with DFT calculations using the regularized-restored strongly constrained and appropriately normed (r$^{2}$SCAN) functional~\cite{furness2020r2scan} on the test set of $4{,}488$ configurations spanning $16$--$128$ $\mathrm{H_2O}$ supercells.
The root-mean-square errors are $0.26~\mathrm{meV}/\mathrm{H_2O}$ for energies and $2.9~\mathrm{meV}/\text{\AA}$ for forces.
Data points fall closely along the diagonal line, and the error distributions shown in the insets are narrow and approximately symmetric, indicating no systematic bias.
This accuracy enables the model to reliably reproduce the meV-scale energy differences between competing proton orderings.

\vspace{1.0em}
\SIheading{Sampling statistics and acceptance rates}

\begin{figure*}[ht]
\centering
\includegraphics[width=1.0\columnwidth]{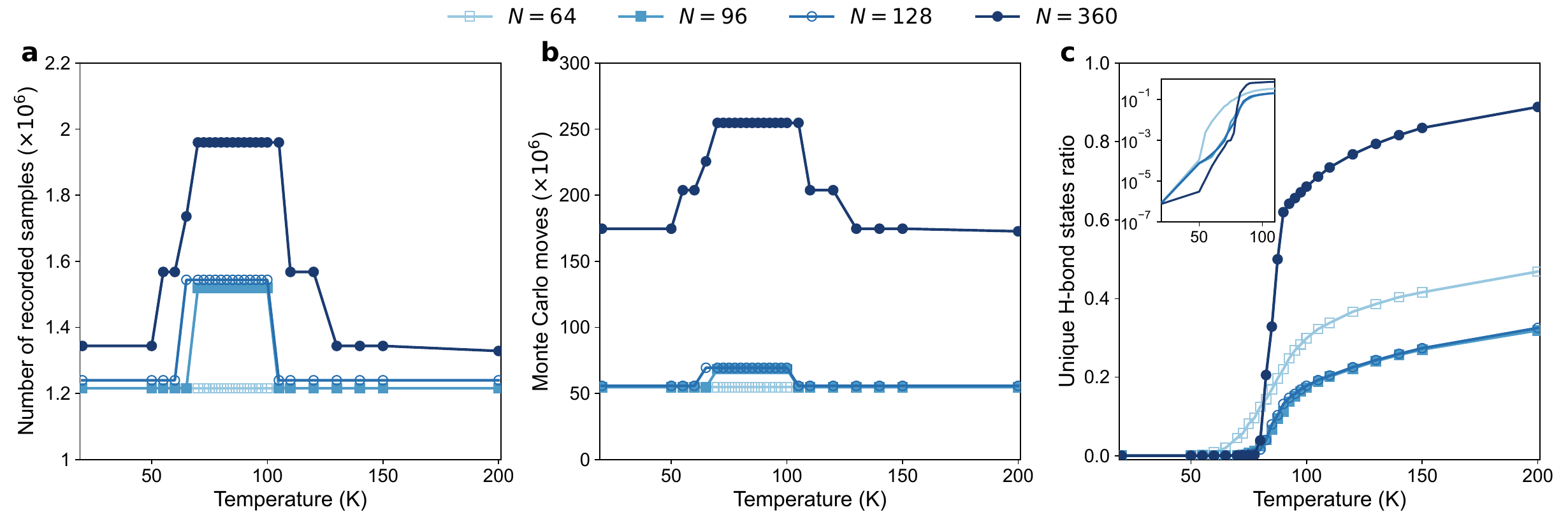}
\caption{
\textbf{Sampling statistics of the Monte Carlo simulations.}
\textbf{a}, Number of recorded post-thermalization samples as a function of temperature.
\textbf{b}, Total number of Monte Carlo move proposals.
\textbf{c}, Ratio of the number of unique visited hydrogen-bond configurations to the total number of recorded samples. The inset shows the same quantity on a logarithmic scale.
}
\label{fig:numsamples}
\end{figure*}

In our composite sampling scheme, one configuration was recorded after each sequence of loop, MALA, and cell updates.
The first 2{,}000 recorded configurations were discarded for thermalization, and the remaining samples were combined across multiple independent Markov chains.
For each system size and temperature, more than $10^6$ retained post-thermalization samples were obtained (Fig.~\ref{fig:numsamples}\textbf{a}) from more than $10^8$ Monte Carlo move proposals (Fig.~\ref{fig:numsamples}\textbf{b}).
Taken together, the full simulation campaign amounted to approximately $3\times10^4$ GPU hours, , measured here in units of NVIDIA RTX 5090 GPU hours.
To characterize configurational exploration, we counted the number of unique hydrogen-bond configurations visited at each temperature and system size and normalized it by the total number of recorded samples.
The resulting ratio, shown in Fig.~\ref{fig:numsamples}\textbf{c}, provides a simple measure of configurational diversity.
Its pronounced suppression upon cooling indicates that the simulations increasingly revisited the same hydrogen-bond configurations at low temperature, particularly for the larger supercells.
Even at $50$--$70~\mathrm{K}$, however, the system still explored hundreds to thousands of distinct hydrogen-bond configurations, consistent with the expectation that configurational diversity is strongly suppressed, but not completely eliminated, in the low-temperature regime.

\begin{figure*}[ht]
\centering
\includegraphics[width=1.0\columnwidth]{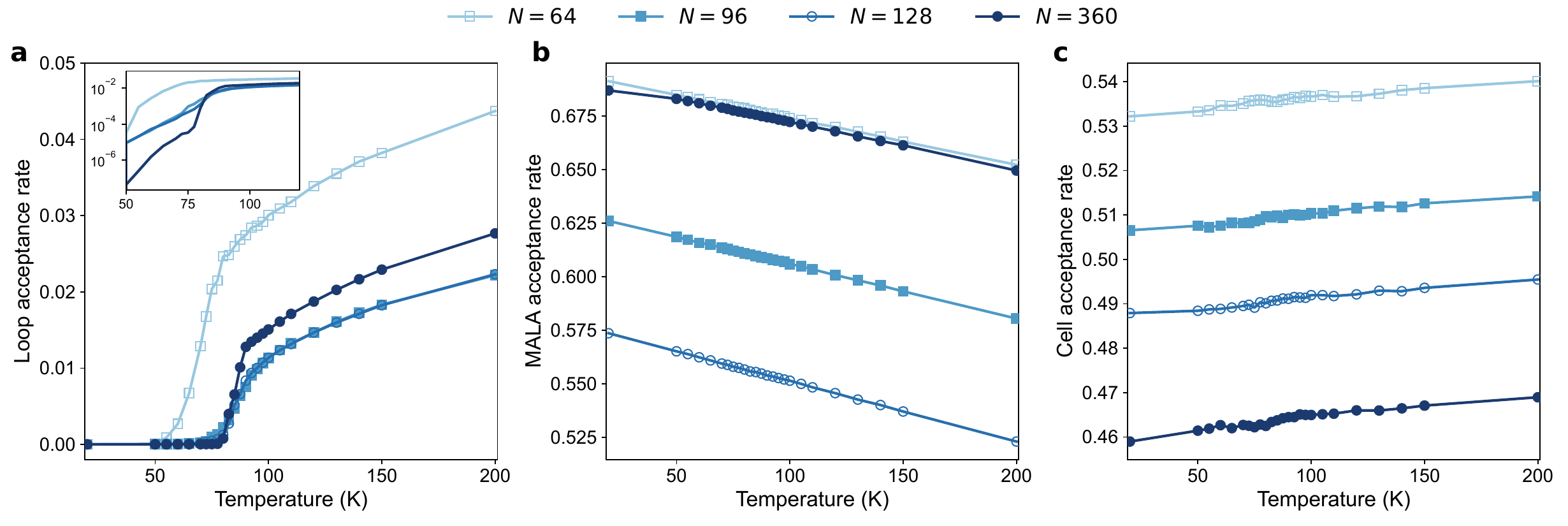}
\caption{
\textbf{Acceptance rates for Monte Carlo updates.}
\textbf{a}, Short-loop updates of the hydrogen-bond configuration. The inset uses a logarithmic scale.
\textbf{b}, Metropolis-adjusted Langevin algorithm updates of the atomic coordinates.
\textbf{c}, Cell-length updates of the orthorhombic simulation cell.
}
\label{fig:acceptance}
\end{figure*}

As shown in Fig.~\ref{fig:acceptance}, the acceptance rates for loop, MALA, and cell updates exhibit distinct temperature dependences.
After tuning the step sizes for each system size and temperature, the MALA and cell acceptance rates remain relatively stable, whereas the loop acceptance rate decreases sharply upon cooling.
For systems with $96$, $128$, and $360$ water molecules, it drops to $0.01\%$ at $80~\mathrm{K}$, while the smallest system with $64$ molecules maintains a substantially higher acceptance rate.
This higher acceptance for the $N=64$ cell is consistent with the absence of clear first-order transition behavior, as the potential energy, heat capacity, and Binder cumulant all vary continuously across the transition region in the main text.
For the larger systems, the strong low-temperature suppression reflects physical freezing into a single proton-ordered configuration, which inhibits topological rearrangements of the hydrogen-bond network.

\vspace{1.0em}
\SIheading{Additional structural and thermodynamic results}

\begin{figure*}[ht]
\centering
\includegraphics[width=1.0\columnwidth]{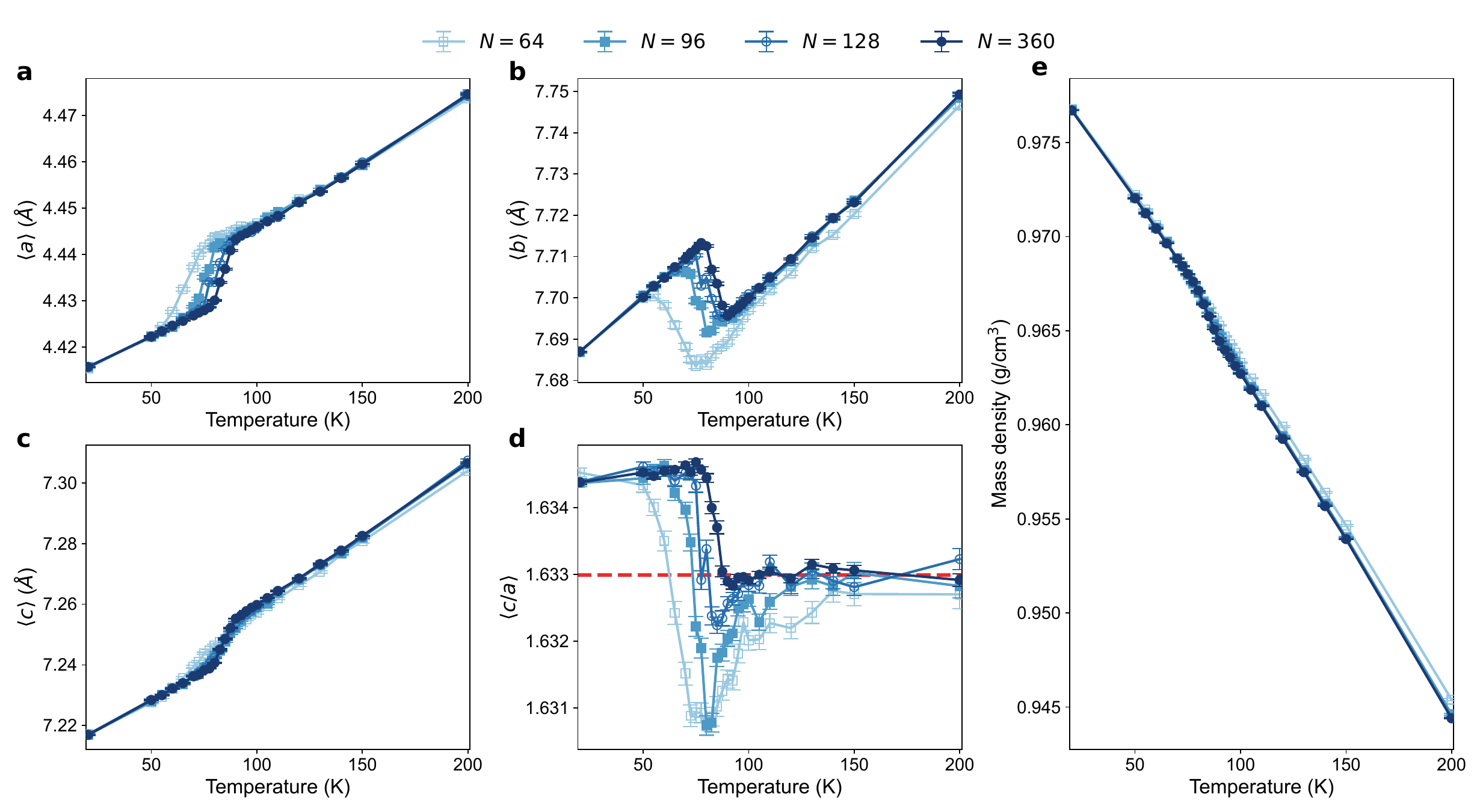}
\caption{
\textbf{Orthorhombic lattice parameters and mass density.}
\textbf{a--c}, Orthorhombic cell lengths $\langle a\rangle$, $\langle b\rangle$, and $\langle c\rangle$ along the three unit-cell axes.
\textbf{d}, Lattice anisotropy ratio $\langle c/a\rangle$. The dashed line indicates the ideal hexagonal close-packed value $\sqrt{8/3}$.
\textbf{e}, Mass density as a function of temperature.
}
\label{fig:cell}
\end{figure*}

As shown in Fig.~\ref{fig:cell}\textbf{a}--\textbf{c}, the orthorhombic lattice parameters exhibit conventional thermal contraction upon cooling from the disordered Ih phase to the ordered XI phase.
The cell lengths $\langle a\rangle$ and $\langle c\rangle$ decrease continuously upon cooling and exhibit a kink near $85$~K, close to the transition temperature inferred from the Binder cumulant and the energy crossing in the main text.
By contrast, $\langle b\rangle$ shows a pronounced anomalous increase near the same temperature, consistent with Ref.~\cite{biggs2024proposing}.
The positions of these kinks shift to higher temperatures as the system size increases.
This size-dependent shift is consistent with the coexistence of ordered and disordered configurations near the transition, where the system samples a multimodal distribution of proton orderings.

As shown in Fig.~\ref{fig:cell}\textbf{d}, the lattice anisotropy ratio $\langle c/a\rangle$ remains close to the ideal hexagonal close-packed value $\sqrt{8/3}$ in the high-temperature disordered ice Ih phase, similar to $\langle b/a\rangle \approx \sqrt{3}$ as discussed in the main text.
In the low-temperature ordered phase, $\langle c/a\rangle$ increases slightly to $1.635$.
This ratio may also depend on the choice of exchange--correlation functional~\cite{rego2020elastic}.
For system sizes $N=64$, $96$, and $128$, $\langle c/a\rangle$ exhibits an anomalous decrease near the transition, but this feature disappears as the system size increases and is absent for $N=360$, indicating a finite-size effect.
Moreover, the temperature at which $\langle c/a\rangle$ rises abruptly shifts to higher values as the system size increases, reaching approximately $85$~K for the largest system with $360$ water molecules.

The mass density shown in Fig.~\ref{fig:cell}\textbf{e} is obtained from the cell volume determined by the lattice parameters.
The density increases upon cooling, consistent with thermal contraction, and exhibits a slight kink near the transition temperature that is more pronounced for larger system sizes.
However, the absolute densities are systematically higher than those measured experimentally~\cite{feistel2006new}.
This bias likely arises from both the equilibrium-structure tendencies inherited from $\mathrm{r^2SCAN}$ and the neglect of nuclear quantum effects, which are known to influence the density of water ice~\cite{herrero2011isotopeiceih}.
Prior benchmarks indicate that SCAN overestimates the density of ice, yielding an overly compact structure~\cite{rego2020elastic}.
As a closely related meta-GGA, $\mathrm{r^2SCAN}$ is expected to exhibit a similar tendency.

\begin{figure*}[ht]
\centering
\includegraphics[width=1.0\columnwidth]{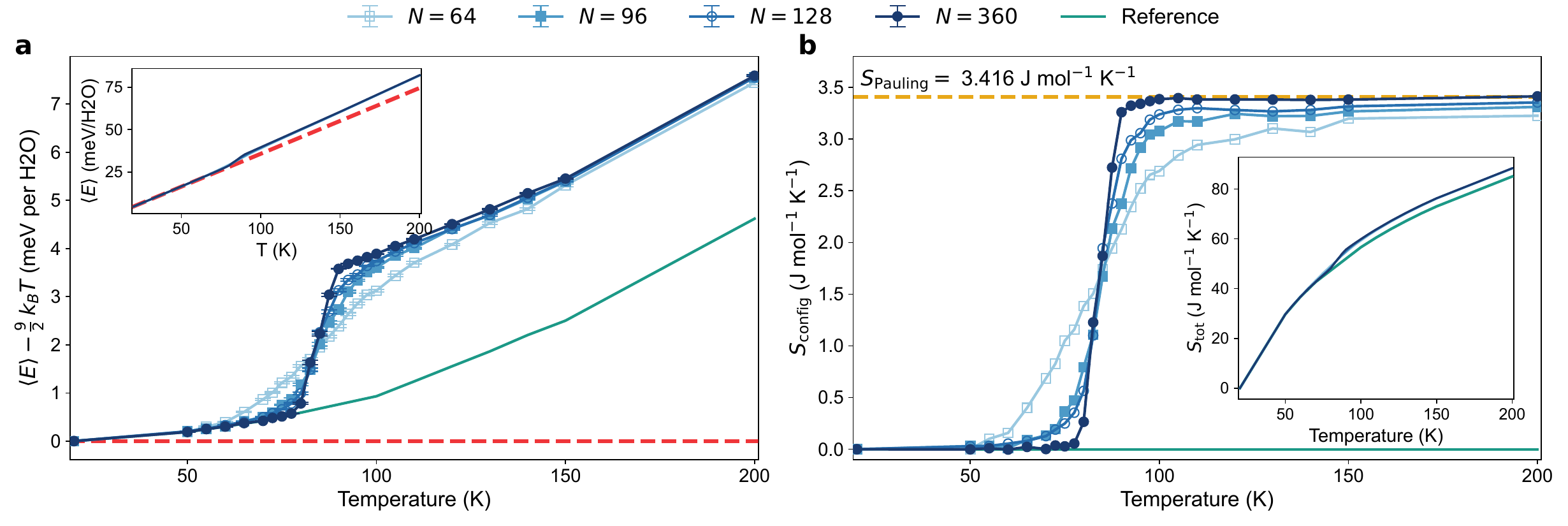}
\caption{
\textbf{Potential energy and entropy across the phase transition.}
\textbf{a}, Potential energy per molecule $\langle E \rangle$ with the classical harmonic contribution $\frac{9}{2}k_{\mathrm{B}}T$ (red dashed line) subtracted, highlighting the anharmonic and hydrogen-bond configurational contributions.
The value labeled ``Reference" defines the entropy baseline.
The inset shows the potential energy without subtraction.
\textbf{b}, Hydrogen-bond configurational entropy $S_{\mathrm{config}}$ obtained by thermodynamic integration.
The green line indicates the reference value defined in panel \textbf{a} and serves as the baseline, whereas the orange dashed horizontal line denotes Pauling's residual entropy.
The inset shows the total entropy $S_{\mathrm{tot}}$ computed from the total potential energy.
}
\label{fig:entropy}
\end{figure*}

Because the potential energy $E$ is dominated by lattice vibrations, we subtract the classical harmonic contribution $\frac{9}{2}k_{\mathrm{B}}T$ to highlight the much smaller energy variations associated with proton ordering. 
The unsubtracted energy is shown in the inset of Fig.~\ref{fig:entropy}\textbf{a}.
To estimate the hydrogen-bond configurational entropy, we construct a reference energy curve. 
From $20$ to $70~\mathrm{K}$, the reference is taken directly from the $N=360$ simulation data. 
From $100$ to $200~\mathrm{K}$, it is obtained by shifting the same data downward by $2.95~\mathrm{meV}/\mathrm{H_2O}$. 
This energy shift is determined independently from simulations at $100~\mathrm{K}$ using an ordered ice XI cell with $360$ water molecules, in which only the continuous degrees of freedom are updated while the hydrogen-bond configuration is kept fixed.
The resulting energy is then compared with the energy from the fully sampled simulation at the same temperature, which includes loop, MALA, and cell updates, to determine the corresponding energy offset.

In Fig.~\ref{fig:entropy}\textbf{b}, the inset shows the total entropy obtained directly from thermodynamic integration of the total potential energy, including both vibrational and hydrogen-bond configurational contributions.
The integration is performed starting from $20~\mathrm{K}$ using $S_{\mathrm{tot}}(T)=\frac{E(T)}{T}-\int_T^{\infty} \frac{E(T^{\prime})}{T^{\prime 2}} dT^{\prime}$~\cite{herrero2013configurational}.
The entropy curves begin to deviate from the reference near $80~\mathrm{K}$, and this difference is identified as the hydrogen-bond configurational contribution $S_{\mathrm{config}}$.
The result after subtraction the reference is shown in the main panel.
For the largest system size, the configurational entropy lost upon proton ordering is close to Pauling's residual entropy~\cite{pauling1935structure, macdowell2004combinatorial, herrero2013configurational, berg2007residual, xu2025equivalence}, indicating that the transition removes nearly all of the configurational entropy associated with proton disorder.
We emphasize that this should be regarded as an estimate, since the separation depends on the construction of the reference curve, and vibrational and configurational degrees of freedom are not strictly separable.

\begin{figure*}[ht]
\centering
\includegraphics[width=1.0\columnwidth]{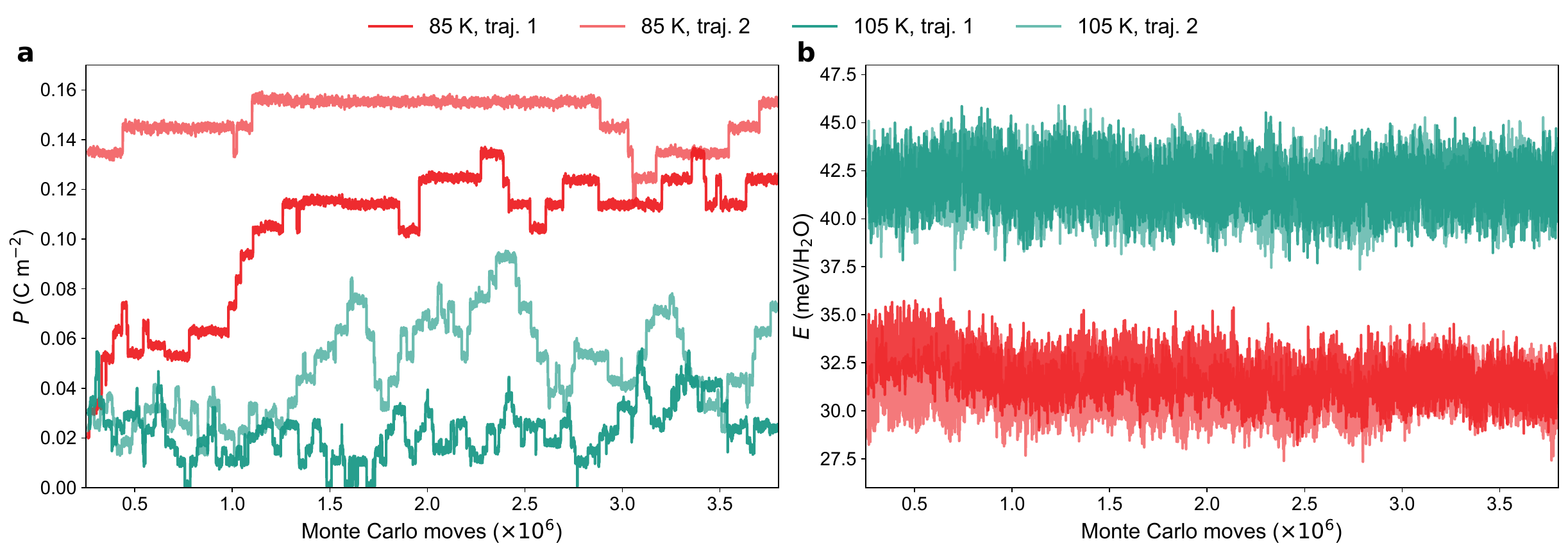}
\caption{
\textbf{Monte Carlo trajectories of polarization and potential energy.}
At each temperature, $85\,\mathrm{K}$ and $105\,\mathrm{K}$, two representative trajectory segments from a 360-water-molecule supercell are shown, randomly selected from two independent runs.
\textbf{a}, Polarization density magnitude $P$.
\textbf{b}, Potential energy per molecule $E$.
}
\label{fig:traj}
\end{figure*}

Figure~\ref{fig:traj} shows two independent Monte Carlo trajectories for an $N=360$ supercell at $85$ and $105\,\mathrm{K}$.
The polarization serves as a structural order parameter, whereas the potential energy reflects the relative thermodynamic stability of the sampled configurations.
In Fig.~\ref{fig:traj}\textbf{a}, the polarization magnitude displays step-like changes because each winding-loop update changes the supercell dipole by a finite increment under periodic boundary conditions, leading to transitions between discrete polarization states.
Small fluctuations around each plateau originate from thermal atomic vibrations.
At $85~\mathrm{K}$, close to the transition temperature, both trajectories occasionally reach the high-polarization state near $0.12\,\mathrm{C\,m^{-2}}$ despite starting from markedly different initial polarizations.
At $105\,\mathrm{K}$, both trajectories explore a broad ensemble of proton-disordered configurations, indicating enhanced configurational sampling and reduced trapping on a single polarization plateau.
Figure~\ref{fig:traj}\textbf{b} shows the corresponding potential energy per molecule.
At $85\,\mathrm{K}$, the trajectory with lower polarization is systematically higher in energy than the trajectory with higher polarization, whereas at $105\,\mathrm{K}$ the two energy traces are nearly indistinguishable.
In addition, for smaller supercells, the trajectories switch between polarization plateaus much more frequently, implying that hydrogen-bond configurational sampling becomes more difficult for larger systems and requires longer trajectories or more independent samples.

\end{document}